\documentclass[journal]{IEEEtran}

\usepackage[T1]{fontenc}
\usepackage[cmex10]{amsmath}
\usepackage{amssymb}
\usepackage{amsthm}
\usepackage{acronym}
\usepackage{threeparttable}
\usepackage{booktabs}
\usepackage[pdftex]{graphicx}
\usepackage{cite}
\usepackage{xcolor}
\usepackage{multirow}
\usepackage{comment}

\usepackage{balance}

\newtheorem{definition}{Definition}
\newtheorem{prop}{Proposition}

\usepackage[caption=false,font=footnotesize]{subfig}
\usepackage{float}
\usepackage{upgreek}
\usepackage{makecell}
\usepackage{mathtools}
\definecolor{blue1}{RGB}{0,60,200}
\usepackage[
unicode=true,
bookmarks=true,
bookmarksnumbered=true,
bookmarksopen=true,
bookmarksopenlevel=1,
hidelinks
]{hyperref}
 
\makeatletter
\let\oldforeign@language\foreign@language
\DeclareRobustCommand{\foreign@language}[1]{%
  \lowercase{\oldforeign@language{#1}}}
\theoremstyle{plain}

\theoremstyle{remark}

\makeatother

\definecolor{blue1}{RGB}{33,92,175}

\newcommand{\bfg}[1]{\boldsymbol{#1}}
\newcommand{\bfb}[1]{\boldsymbol{\rm #1}}
\newcommand{\T}{^{\intercal}}

\newcommand{\nx}{n}
\newcommand{\ny}{m}
\newcommand{\neta}{r}

\newcommand{\jac}[2]{\bfg{{#1}}_{\hspace{-0.2mm}#2}}

\newcommand{\reigvmat}{ \bfg \Phi}
\newcommand{\leigvmat}{ \bfg \Psi}
\usepackage{float}

\newcommand{\reigv}{\bfg \phi}
\newcommand{\leigv}{\bfg \psi}

\newcommand{\Contr}{\bfb{\mathcal{P}}}
\newcommand{\contrel}{\mathcal{P}}
\newcommand{\Contra}{\bfb{\mathcal{C}}}
\newcommand{\contrael}{\mathcal{C}}

\newcommand{\mresp}{\bfg \xi}

\acrodef{avr}[AVR]{automatic voltage regulator}
\acrodef{pss}[PSS]{power system stabilizer}
\acrodef{ode}[ODE]{ordinary differential equation}
\acrodef{dae}[DAE]{differential algebraic equation}
\acrodef{sdae}[SDAE]{stochastic differential algebraic equation}
\acrodef{sm}[SG]{synchronous generator}
\acrodef{sm}[SM]{synchronous machine}
\acrodef{tds}[TDS]{time-domain simulations}
\acrodef{tg}[TG]{turbine governor}
\acrodef{der}[DER]{distributed energy resource}
\acrodef{agc}[AGC]{automatic generation control} 
\acrodef{pll}[PLL]{phase-locked loop} 
\acrodef{tdi}[TDI]{Time Domain Integration}
\acrodef{aiits}[AIITS]{all-island Irish transmission system}
\acrodef{pf}[PF]{participation factor}
\acrodef{sssa}[SSSA]{small-signal stability analysis}
\acrodef{mimo}[MIMO]{Multiple-Input Multiple-Output}
\acrodef{cf}[CF]{complex frequency}
\acrodef{sma}[SMA]{selective modal analysis}
\acrodef{coi}[COI]{center-of-inertia}
\acrodef{nf}[NF]{normal form}
\acrodef{vsm}[VSM]{virtual synchronous machine}
\acrodef{fr}[FR]{frequency regulation}
\acrodef{scr}[SCR]{short-circuit ratio}
\acrodef{scc}[SCC]{short-circuit capacity}
\acrodef{ibr}[IBR]{inverter-based resources}
\acrodef{gfm}[GFM]{grid-forming}
\acrodef{gfl}[GFL]{grid-following}
\acrodef{cdf}[CDF]{cumulative distribution function}
\acrodef{mc}[MC]{Monte Carlo} 

\title{Mode-in-Variance Participation Factors under Stochastic Excitation}

\author{Bruno Pinheiro, \textit{Member, IEEE}, 
Liya Huang, \textit{Student Member, IEEE}, \\ 
and Georgios Tzounas, 
\textit{Senior Member, IEEE}
\thanks{The authors are with the School of Electrical and Electronic Engineering, University College Dublin, Dublin, D04V1W8, Ireland. \\
Corresponding author's e-mail: \protect\href{http://georgios.tzounas@ucd.ie}{georgios.tzounas@ucd.ie}.

This work is supported by the CETPartnership Joint Call 2024 under project NU-ACTIS, co-funded by the European Commission (grant no. 101069750). The participation of University College Dublin is funded by the Sustainable Energy Authority of Ireland (ID 24/RDD/1390).
}
}
    
\begin{document}

\maketitle
\pagestyle{plain} 
\IEEEpeerreviewmaketitle
 
\begin{abstract}
This paper investigates how the dynamic modes of a power system contribute to the stationary variance of its variables under continuous stochastic excitation. Using a modal decomposition of the covariance matrix, the variance of each state or algebraic variable can be resolved into components associated with individual modal responses and their interactions. These components are shown to correspond to within-mode variance and cross-modal covariance, respectively, and the corresponding normalized contributions are termed \textit{mode-in-variance} \acp{pf}.  The proposed \acp{pf} can be used to identify modes governing the variability of important system quantities such as frequency and voltage, quantify the effect of modal interactions, and reveal control dynamics that can be targeted to reduce their stochastic variability. These capabilities are demonstrated through case studies based on the Kundur two-area and IEEE 39-bus systems.
\end{abstract}

\section{Introduction}

\subsection{Motivation}

\IEEEPARstart{P}{ower} systems are continuously subjected to stochastic forcing arising from fluctuations in loads and primary energy resources, such as wind and solar generation \cite{bendong2026, bruno_sin}.  Such forcing induces variability in important system quantities, including frequency and voltage \cite{Yuqi2026,taulant_2026,Samuel2019}. Existing methods can quantify the magnitude of this variability, but they do not explain which dynamic modes are responsible for it or how different modal contributions combine in the observed response. Identifying the dynamic modes contributing to stochastic variability can provide a means to characterize its origin and support control design. This motivates the modal perspective adopted in this work.

\subsection{Literature Review}

Power systems subject to continuous stochastic variations can be represented through \acp{sdae} \cite{federico2013,milano2022power}.  Existing works have adopted both numerical and analytical approaches within this framework \cite{georgia_2020,brunogm_2026,Hongyu2019,Hafsah2025,adeen2023,vorobev_2019,Goodarz2016}. For example, Monte Carlo time-domain simulations have been used to study voltage and frequency dynamics under stochastic excitation \cite{brunogm_2026,Hafsah2025}. Analytical studies have examined statistical properties of the system response, including stationary variance and frequency distributions \cite{Hongyu2019,adeen2023,vorobev_2019,Goodarz2016}.

A systematic basis for relating system variables to dynamic modes is provided by classical modal analysis \cite{tzounas2019modal}.
Probabilistic small-signal  approaches have extended modal analysis to describe the distributions of eigenvalues, damping ratios, and participation factors induced by uncertain operating conditions \cite{rev_milanovic2019,luke2026,rongcan2026,CYChung_2018,CYChung2012}. Related studies have examined modal variations across prescribed ranges of operating conditions and parameters \cite{luke_SEGN,murilo2020,interval2022}. 

A complementary line of work has examined how the modal structure of a system is reflected in energy-based response measures following initial perturbations.  In particular, \cite{iskakov2021,iskakov2023} propose Lyapunov modal analysis to attribute accumulated state or signal energy to individual modes and their interactions. Modal attribution following initial perturbations has also been used to analyze the magnitude of voltage and frequency deviations.  In particular, \cite{benedetti2024modal} defines small-signal variability in terms of the maximum linearized time-domain deviation and quantifies the contribution of individual modes from the corresponding decoupled modal responses.

To the best of our knowledge, a systematic approach to linking stochastic fluctuations in power system quantities arising from continuous variations in demand and generation to the underlying dynamic modes of the system is currently missing.  Such an approach is particularly relevant because the observed stochastic variability cannot be attributed solely to the characteristics of the stochastic inputs, as the system dynamics also shape how these fluctuations propagate to individual system variables.
Moreover, the modes that dominate stochastic variability need not coincide with those identified as most important by classical modal participation factors.  This paper aims to address this gap.

\subsection{Contribution}

The contribution of this paper is twofold, as follows:
\begin{itemize}

\item A modal decomposition of the stationary covariance matrix is applied to define \textit{mode-in-variance} \acp{pf}, which quantify the contribution of dynamic modes to the stationary variance of system variables under continuous stochastic excitation. The approach is formulated for both state and algebraic variables, enabling modal contributions to the variability of quantities such as frequency and voltage to be studied.

\item The contribution of each mode is resolved into a direct component associated with within-mode variance, an interaction component associated with cross-modal covariance, and a net component combining both effects. The interaction and net contributions are signed, thereby accounting for reinforcement or cancellation among modal responses.
\end{itemize}

\subsection{Paper Organization}

The remainder of this paper is organized as follows. 
Section~\ref{sec:model} provides background on the modeling of power systems under stochastic variations and defines the stationary covariance matrix of the system under small stochastic perturbations.
Section~\ref{sec:modalP} introduces the proposed mode-in-variance \acp{pf} and presents their formulation for both state and algebraic variables. Section~\ref{sec:case_studies} evaluates the proposed approach through case studies based on the Kundur two-area and IEEE 39-bus systems. Finally, Section~\ref{sec:conclu} draws conclusions and discusses directions for future work.

\section{Preliminaries}
\label{sec:model}

\subsection{Stochastic Power System Model}
\label{sec:sdae}

Power systems subject to stochastic variations can be modeled by a set of nonlinear \acp{sdae} \cite{federico2013}, as follows:
\begin{equation}
\label{eq:sdae}
\begin{aligned}
\dot{\bfg x} &= \bfg f(\bfg x,\bfg y,\bfg\eta) \, , \\
\bfg 0 &= \bfg g(\bfg x,\bfg y,\bfg\eta) \, , 
  \end{aligned}
\end{equation}
where $\bfg x= \bfg x(t) \in\mathbb{R}^{\nx}$ and $\bfg y= \bfg y(t)\in\mathbb{R}^{\ny}$ are the state and algebraic variables, respectively; $\bfg f : \mathbb{R}^{\nx+\ny+\neta} \rightarrow \mathbb{R}^{\nx}$ and $\bfg g : \mathbb{R}^{\nx+\ny+\neta} \rightarrow \mathbb{R}^{\ny}$ define, respectively, the differential and algebraic equations; 
$\bfg\eta=\bfg\eta(t)\in\mathbb{R}^{\neta}$ is the vector of stochastic inputs representing fluctuations in generation and demand, entering the system as continuous external forcing. The dynamics of 
$\bfg\eta$ are represented as a multivariate It\^o diffusion process
\begin{align}
\label{eq:diffusion}
d\bfg\eta &= \bfg a(\bfg\eta)\,dt
+ \bfb B(\bfg\eta)\,d\bfg w \, , 
\end{align}
where $\bfg a$ is the drift vector, which describes the deterministic evolution of $\bfg\eta$; $\bfb B$ is the diffusion matrix, which determines the intensity of the random fluctuations; and $\bfg w$ is a vector of independent Wiener processes.
The stochastic response of the nonlinear model in \eqref{eq:sdae}--\eqref{eq:diffusion} can be studied through time-domain simulations.
In this work, the stochastic inputs are represented by the commonly used multivariate Ornstein--Uhlenbeck (OU) process, obtained from \eqref{eq:diffusion} by considering a linear mean-reverting drift and a constant diffusion matrix:
\begin{equation}
\label{eq:oup}
d\bfg\eta
=
-\bfb A_\eta(\bfg\eta-\bfg\mu)\,dt
+
\bfb B_\eta\,d\bfg w \; ,
\end{equation}
where $\bfg\mu$ is the vector of mean values, $\bfb A_\eta$ is the mean-reversion matrix, and $\bfb B_\eta$ is the diffusion matrix.

\subsection{Small Disturbance Analysis}
\label{sec:sssa}

For sufficiently small disturbances, \eqref{eq:sdae} can be linearized around an equilibrium point $(\bfg x_0,\bfg y_0,\bfg\eta_0)$, with $\bfg\eta_0=\bfg\mu$, allowing the stochastic response to be characterized analytically.
Under the standard assumption that $\jac{g}{y}$ is nonsingular, the algebraic variables can be locally eliminated. Combining the resulting linearized model with \eqref{eq:oup} gives
\begin{align}
\nonumber
d\bfg z
&=
\begin{bmatrix}
\jac{f}{x}
-
\jac{f}{y}
\jac{g}{y}^{-1}
\jac{g}{x} & \jac{f}{\eta}
-
\jac{f}{y}
\jac{g}{y}^{-1}
\jac{g}{\eta} \\
\bfg 0 & -\bfb A_\eta
\end{bmatrix}
\bfg z dt +
\begin{bmatrix}
\bfg 0 \\
\bfb B_\eta
\end{bmatrix}
d\bfg w
\\
&=
\bfb A_0\bfg z dt+\bfb B_0 d\bfg w  \; ,
\label{eq:linear_sdae}
\end{align}
where $\bfg z = [\Delta\bfg x^T , 
\Delta\bfg\eta^T]^T$, 
with
$\Delta\bfg x=\bfg x-\bfg x_0$ and
$\Delta\bfg\eta=\bfg\eta-\bfg\mu$,
and $\jac{f}{x}, \jac{f}{y}, \jac{f}{\eta}, \jac{g}{x}, \jac{g}{y}, \jac{g}{\eta}$ are the Jacobians evaluated at the equilibrium point.
The solution of \eqref{eq:linear_sdae} is given by
\begin{equation}
\label{eq:analy_solu}
\bfg z(t) =
e^{\bfb A_0t}\bfg z(0) +
\int_0^t e^{\bfb A_0(t-\tau)}
\bfb B_0\,d\bfg w(\tau) \; .
\end{equation}

The first term represents the deterministic response to the initial condition $\bfg z(0)$, while the second term represents the response to the stochastic forcing. The expectation of the stochastic integral in \eqref{eq:analy_solu} is zero \cite{kloeden1999numerical}, such that 
\begin{align}
\label{eq:Ez}
\mathbb{E}[\bfg z(t)]
= e^{\bfb A_0t}\bfg z(0) \, ,
\end{align}
where $\mathbb{E}[\cdot]$ denotes the expectation operator.  We thus have
\begin{equation}
\label{eq:mean_solution}
\bfg z(t)-\mathbb{E}[\bfg z(t)]
=
\int_0^t
e^{\bfb A_0(t-\tau)}
\bfb B_0\,d\bfg w(\tau) \; .
\end{equation}

Since the right-hand side is an integral of Wiener increments, $\bfg z(t)$ is Gaussian. Moreover, if all eigenvalues of $\bfb A_0$ have negative real parts (which is equivalent to both $\jac{f}{x}-\jac{f}{y}\jac{g}{y}^{-1}\jac{g}{x}$ and $-\bfb A_\eta$ being Hurwitz),
then
$\lim_{t\rightarrow\infty}\mathbb{E}[\bfg z(t)]=0$ from \eqref{eq:Ez}, i.e., the stationary distribution of $\bfg z$ has zero mean.

The stochastic variability of $\bfg z(t)$ can be characterized through its covariance matrix, defined as
\begin{equation}
\label{eq:cov_definition}
\bfb P(t)
=
\mathbb{E}
\left[
\left(
\bfg z(t)-\mathbb{E}[\bfg z(t)]
\right)
\left(
\bfg z(t)-\mathbb{E}[\bfg z(t)]
\right)^T
\right] \; .
\end{equation}

Substituting \eqref{eq:mean_solution} into \eqref{eq:cov_definition} and
applying the It\^{o} isometry property \cite{kloeden1999numerical}
gives
\begin{align}
\bfb P(t)
=
\int_0^t
e^{\bfb A_0(t-\tau)}
\bfb Q
e^{\bfb A_0^T(t-\tau)}
\,d\tau  \; ,
\end{align}
where
\begin{equation}
\label{eq:Q}
\bfb Q=\bfb B_0\bfb B_0^T \; .
\end{equation}

Taking $t\rightarrow\infty$ and applying the change of variable $s = t- \tau$ gives the stationary covariance matrix
\begin{equation}
\label{eq:stationary_cov}
\bfb P = 
\lim_{t \rightarrow \infty }
\bfb P(t)
=
\int_0^\infty
e^{\bfb A_0\tau}
\bfb Q
e^{\bfb A_0^T\tau}
\,d\tau  \; .
\end{equation}

Equation \eqref{eq:stationary_cov} shows that the stationary covariance results from the accumulation of stochastic disturbances propagated through the system dynamics. Hence, it is jointly determined by the stochastic forcing, represented by $\bfb Q$, and the system dynamics, represented by the state-transition matrix $e^{\bfb A_0\tau}$. Observe that, since $\bfb Q=\bfb B_0\bfb B_0^T$, \eqref{eq:stationary_cov} is also the controllability Gramian associated with \eqref{eq:linear_sdae}.

The diagonal elements of $\bfb P$ correspond to the stationary variances of the variables in $\bfg z$, i.e., for $z_k$:
\begin{equation}
\label{eq:state_variance}
P_k \triangleq \bfb P(k,k) = 
\operatorname{Var}(z_k)= \sigma^2_{z_k} \; ,
\end{equation}
while the off-diagonal elements quantify the covariance between pairs of variables.

Equivalently to \eqref{eq:stationary_cov}, the stationary covariance matrix satisfies the Lyapunov equation
%
\begin{equation}
\label{eq:lyapunov}
\bfb A_0\bfb P
+ \bfb P\bfb A_0^T
+ \bfb Q = 
\bfb 0 \; .
\end{equation}
The derivation of \eqref{eq:lyapunov} is provided in the Appendix.

\section{Modal Characterization of \\Stochastic Variability}
\label{sec:modalP}

This section presents the proposed approach for quantifying the contribution of system dynamic modes to the stationary variance of state and algebraic variables.

\subsection{Spectral Decomposition of the Covariance Matrix}
\label{sec:state_spf}

The stationary covariance in \eqref{eq:stationary_cov} can be decomposed in terms of the eigenvalues and eigenvectors of $\bfb A_0$. Let ${\reigvmat} = [{\reigv}_1 \ \cdots \ {\reigv}_\nu]$ and ${\leigvmat} = [{\leigv}_1\T \ \cdots \ {\leigv}_\nu\T]\T$, where the columns of ${\reigvmat}$ are the right eigenvectors of $\bfb A_0$, and the rows of ${\leigvmat}$ are the corresponding left eigenvectors; $\nu=\nx+\neta$. 
Under the assumption that $\bfb A_0$ is diagonalizable, the left and right eigenvectors are normalized such that ${\leigvmat} {\reigvmat} = \bfb I$.   
Accordingly, we can write $\bfb A_0={\reigvmat}\bfg\Lambda{\leigvmat}$,
where
$
\label{eq:modal_matrix}
\bfg \Lambda
=
\operatorname{diag}
\left(
\lambda_1,\ldots,\lambda_\nu
\right)
$
contains the eigenvalues of $\bfb A_0$. The state-transition matrix $e^{\bfb A_0\tau}$ can then be expressed as
\begin{equation}
\label{eq:modal_transition}
e^{\bfb A_0\tau}
=
{\reigvmat} e^{\bfg\Lambda\tau}{\leigvmat}
=
\sum_{i=1}^{\nu}
e^{\lambda_i\tau}
{\reigv}_i{\leigv}_i \; .
\end{equation}

Substitution of \eqref{eq:modal_transition} into \eqref{eq:stationary_cov} gives
\begin{equation}
\label{eq:expansion}
\begin{aligned}
\bfb P
&=
\int_0^\infty
\left(
\sum_{i=1}^{\nu} e^{\lambda_i\tau}
{\reigv}_i{\leigv}_i
\right)
\bfb Q
\left(
\sum_{j=1}^{\nu} e^{\lambda_j\tau}
{\leigv}_j^T{\reigv}_j^T
\right)
d\tau
\\
&=
\sum_{i=1}^{\nu}\sum_{j=1}^{\nu}
{\reigv}_i
\left(
{\leigv}_i\bfb Q{\leigv}_j^T
\right)
{\reigv}_j^T
\int_0^\infty
e^{(\lambda_i+\lambda_j)\tau}
d\tau  \; .
\end{aligned}
\end{equation}

For asymptotically stable $\bfb A_0$, $\operatorname{Re}(\lambda_i+\lambda_j)<0$, and
\begin{equation}
\label{eq:modal_integral}
\int_0^\infty
e^{(\lambda_i+\lambda_j)\tau}
d\tau
=
-\frac{1}{\lambda_i+\lambda_j} \; .
\end{equation}

It follows that matrix $\bfb P$ can be written as
\begin{equation}
\label{eq:exact_modal_covariance}
\bfb P
=
-\sum_{i=1}^{\nu}\sum_{j=1}^{\nu}
\frac{
{\reigv}_i
\left(
{\leigv}_i\bfb Q{\leigv}_j^T
\right)
{\reigv}_j^T
}{
\lambda_i+\lambda_j
} \; .
\end{equation}
which expresses $\bfb P$ as a sum of contributions associated with pairs of eigenvalues. This pairwise structure arises because covariance is quadratic in the system response (see \eqref{eq:cov_definition}). In 
\eqref{eq:expansion}, the modal expansion therefore enters 
on both sides of $\bfb Q$, producing one modal index from each side and, hence, contributions associated with pairs $(i,j)$.
Accordingly, the contribution associated with the eigenvalue pair $(i,j)$ is
\begin{equation}
\label{eq:modal_pair_covariance}
\Contr_{ij}
=
-\frac{
{\reigv}_i
\left(
{\leigv}_i\bfb Q{\leigv}_j^T
\right)
{\reigv}_j^T
}{\lambda_i+\lambda_j} \; , 
\qquad {\rm with} \ \ 
\Contr_{ij}\in\mathbb{C}^{\nu\times \nu} \ .
\end{equation}

Matrix $\bfb P$ is then recovered as
\begin{equation}
\label{eq:cov_sum_pairs}
\bfb P =
\sum_{i=1}^{\nu}\sum_{j=1}^{\nu}
\Contr_{ij} \; .
\end{equation}

We note that an analogous spectral decomposition in terms of pairs of eigenvalues can be obtained in a deterministic free response analysis for the accumulated response energy following an initial perturbation \cite{iskakov2021}.  However, the objective and quantity of interest in the present work are different, as the decomposition in \eqref{eq:cov_sum_pairs} serves as the basis for identifying modal contributions to the stationary variance of system variables.
To establish the statistical meaning of these contributions, 
we substitute the modal representation of $e^{\bfb A_0\tau}$ in \eqref{eq:modal_transition} into the centered stochastic response in \eqref{eq:mean_solution}. This gives
\begin{equation}
\label{eq:modal_response_sum}
\bfg z(t)-\mathbb{E}[\bfg z(t)]
=
\sum_{i=1}^{\nu}\mresp_i(t) \; ,
\end{equation}
where 
\begin{equation}
\label{eq:modal_stochastic_response}
\mresp_i(t)
= \int_0^t
e^{\lambda_i(t-\tau)}
{\reigv}_i{\leigv}_i
\bfb B_0\,d\bfg w(\tau) \; ,
\end{equation}
denotes the contribution of the $i$th eigenvalue to the centered stochastic response of $\bfg z(t)$.

\begin{prop}
\label{prop:modal_cross_covariance}
The quantity $\Contr_{ij}$
is equal to
the stationary nonconjugate cross-covariance between the contributions of the $i$th and $j$th eigenvalues to the stochastic response, i.e.,
\begin{equation}
\label{eq:modal_cross_covariance}
\lim_{t\rightarrow\infty}
\mathbb{E}
\left[
\mresp_i(t)\mresp_j^T(t)
\right]
= \Contr_{ij} \; .
\end{equation}
\end{prop}

The proof of \eqref{eq:modal_cross_covariance} is given in the Appendix.

Each complex-conjugate eigenvalue pair is next grouped into a single oscillatory mode.  Let $\mathcal{M}_h$ denote the set of eigenvalue indices associated with the $h$th mode. Consequently, each $\mathcal{M}_h$
contains either one index, corresponding to a 
real eigenvalue representing a nonoscillatory mode, or two indices, corresponding to a complex-conjugate pair. 
The contribution of each mode $\mathcal{M}_h$ to the stochastic response is defined as
\begin{equation}
\label{eq:physical_modal_response}
\mresp_{\scriptscriptstyle\mathcal{M}_h}(t)
=
\sum_{i\in\mathcal{M}_h}
\mresp_i(t) \; .
\end{equation}

From \eqref{eq:modal_response_sum} and
\eqref{eq:physical_modal_response}, the centered stochastic response can then be written as
\begin{equation}
\bfg z(t)-\mathbb{E}[\bfg z(t)]
=
\sum_h \mresp_{\scriptscriptstyle\mathcal{M}_h}(t) \; .
\end{equation}

Substituting into \eqref{eq:cov_definition}, separating within-mode from cross-mode terms, 
and taking the limit $t\rightarrow\infty$,
the stationary covariance can be written as
\begin{equation}
\label{eq:physical_mode_covariance_decomposition}
\begin{aligned}
\bfb P ={}&
\lim_{t\rightarrow\infty}
\sum_h
\mathbb{E}
\left[
\mresp_{\scriptscriptstyle\mathcal{M}_h}
\mresp_{\scriptscriptstyle\mathcal{M}_h}^T
\right]
\\ &
+
\lim_{t\rightarrow\infty}
\sum_{h<\ell} \mathbb{E}
\left[
\mresp_{\scriptscriptstyle\mathcal{M}_h}
\mresp_{\scriptscriptstyle\mathcal{M}_\ell}^T
+
\mresp_{\scriptscriptstyle\mathcal{M}_\ell}
\mresp_{\scriptscriptstyle\mathcal{M}_h}^T
\right]
\, .
\end{aligned}
\end{equation}

The first term represents the stationary covariance associated with individual modes, whereas the second term represents the covariance between distinct modes.

\subsection{Mode-in-Variance Participation Factors}
\label{sec:spf}

For variable $z_k$, with stationary variance $P_k$, the contribution of pair $(i,j)$ is obtained from \eqref{eq:modal_pair_covariance} as
\begin{equation}
\label{eq:state_modal_pair_variance}
\contrel_k^{(i,j)}
\triangleq
\bfg e_k^T\Contr_{ij}\bfg e_k
=-\frac{\left(\bfg e_k^T{\reigv}_i\right)
\left({\leigv}_i\bfb Q{\leigv}_j^T\right)
\left({\reigv}_j^T\bfg e_k\right)
}{\lambda_i+\lambda_j} 
\; ,
\end{equation}
where $\bfg e_k$ is the $k$th canonical basis vector in $\mathbb{R}^{\nu}$.
Hence,
\begin{equation}
\label{eq:state_variance_modal_decomposition}
P_k = \sum_{i=1}^{\nu}\sum_{j=1}^{\nu}
\contrel_k^{(i,j)} \; .
\end{equation}

The contributions of the eigenvalue pairs to $P_k$ are subsequently grouped into \textit{direct} and \textit{interaction} contributions.

The direct contribution associated with a single mode $\mathcal{M}_h$ is obtained by collecting the terms with $i,j \in \mathcal{M}_h$:
\begin{equation}
\label{eq:within_mode_variance}
\boxed{
P_k^{[\mathcal{M}_h,\mathrm{dir}]}
\triangleq
\sum_{i\in\mathcal{M}_h}
\sum_{j\in\mathcal{M}_h}
\contrel_k^{(i,j)} \; .
}
\end{equation}

The pairwise interaction contribution between two distinct modes $\mathcal{M}_h$ and $\mathcal{M}_\ell$ is obtained by collecting the corresponding cross mode terms:
\begin{equation}
\label{eq:between_mode_variance}
P_k^{[\mathcal{M}_h,\mathcal{M}_\ell]}
\triangleq
\sum_{i\in\mathcal{M}_h}
\sum_{j\in\mathcal{M}_\ell}
\left(\contrel_k^{(i,j)} +
\contrel_k^{(j,i)} \right),
\quad h\neq \ell \; .
\end{equation}

Summing $P_k^{[\mathcal{M}_h,\mathcal{M}_\ell]}$ over all $h\neq \ell$, we define the total interaction contribution of mode $\mathcal{M}_h$ as
\begin{equation}
\label{eq:mode_interaction_variance}
\boxed{
P_k^{[\mathcal{M}_h,\mathrm{int}]}
\triangleq
\frac{1}{2}
\sum_{\ell\neq h}
P_k^{[\mathcal{M}_h,\mathcal{M}_\ell]} \; ,}
\end{equation}
where the factor $1/2$ is introduced because 
each pairwise contribution between modes
$\mathcal{M}_h$ and $\mathcal{M}_\ell$ enters the interaction term of both modes. To prevent the same quantity from being counted twice when the modal contributions are summed, one half is assigned to each mode.

From \eqref{eq:within_mode_variance} and \eqref{eq:mode_interaction_variance}, the stationary variance in \eqref{eq:state_variance_modal_decomposition} can be written as
\begin{equation}
\label{eq:variance_mode_decomposition}
P_k
=
\sum_h
\left[
P_k^{[\mathcal{M}_h,\mathrm{dir}]}
+
P_k^{[\mathcal{M}_h,\mathrm{int}]}
\right] \; .
\end{equation}

\begin{prop}
\label{prop:2}
The direct contribution $P_k^{[\mathcal{M}_h,\mathrm{dir}]}$ represents the part of the stationary variance  of $z_k$ associated with mode $\mathcal{M}_h$ and is therefore nonnegative. 
Moreover, the interaction contribution $P_k^{[\mathcal{M}_h,\mathrm{int}]}$ represents the part of the stationary variance of $z_k$ arising from the cross covariances between mode $\mathcal{M}_h$ and the remaining modes, and can therefore be positive or negative.
\end{prop}

\begin{proof}
Using \eqref{eq:physical_mode_covariance_decomposition},  $P_k^{[\mathcal{M}_h,\mathrm{dir}]}$ can be expressed as
\begin{equation}
\begin{aligned}
P_k^{[\mathcal{M}_h,\mathrm{dir}]}
&= \lim_{t\rightarrow\infty}
\bfg e_k^T
\mathbb{E}
\left[
\mresp_{\scriptscriptstyle\mathcal{M}_h}
\mresp_{\scriptscriptstyle\mathcal{M}_h}^T
\right]
\bfg e_k 
\\
&=
\lim_{t\rightarrow\infty}
\mathbb{E}
\left[
\left(
\bfg e_k^T
\mresp_{\scriptscriptstyle\mathcal{M}_h}
\right)^2
\right]
\geq 0 \; .
\end{aligned}
\end{equation}

Moreover, for two distinct modes $\mathcal{M}_h$ and $\mathcal{M}_\ell$,
the pairwise interaction contribution is
\begin{equation}
\begin{aligned}
P_k^{[\mathcal{M}_h,\mathcal{M}_\ell]}
&= \lim_{t\rightarrow\infty}
\bfg e_k^T
\mathbb{E}
\left[
\mresp_{\scriptscriptstyle\mathcal{M}_h}
\mresp_{\scriptscriptstyle\mathcal{M}_\ell}^T
+
\mresp_{\scriptscriptstyle\mathcal{M}_\ell}
\mresp_{\scriptscriptstyle\mathcal{M}_h}^T
\right]
\bfg e_k
\\
&=
2\lim_{t\rightarrow\infty}
\,\operatorname{Cov}
\left(
\bfg e_k^T\mresp_{\scriptscriptstyle\mathcal{M}_h},
\bfg e_k^T\mresp_{\scriptscriptstyle\mathcal{M}_\ell}
\right) 
\, .
\end{aligned}
\end{equation}
Using \eqref{eq:mode_interaction_variance}, it follows that
\begin{equation}
P_k^{[\mathcal{M}_h,\mathrm{int}]}
=
\sum_{\ell\neq h}
\lim_{t\rightarrow\infty}
\operatorname{Cov}
\left(
\bfg e_k^T\mresp_{\scriptscriptstyle\mathcal{M}_h},
\bfg e_k^T\mresp_{\scriptscriptstyle\mathcal{M}_\ell}
\right) \; .
\end{equation}
Since covariances can be positive or negative, $P_k^{[\mathcal{M}_h,\mathrm{int}]}$ is a signed quantity.
\end{proof}

\vspace{1mm}

A positive (negative) $P_k^{[\mathcal{M}_h,\mathrm{int}]}$ indicates reinforcement 
(cancellation) 
between the contribution of $\mathcal{M}_h$ to $z_k$ and the contributions of the remaining modes.

\begin{definition}[Mode-in-variance PFs]
Based on the variance decomposition in \eqref{eq:variance_mode_decomposition}, the direct and interaction contributions of mode $\mathcal{M}_h$ to the
stationary variance of $z_k$ are normalized by $P_k$. 
The corresponding
normalized quantities are defined as the direct and interaction
\textit{mode-in-variance} \acp{pf}:
\begin{equation*}
\pi_{k,h}^{\mathrm{dir}}
\triangleq
\frac{
P_k^{[\mathcal{M}_h,\mathrm{dir}]}
}{
P_k
} \ , \qquad 
\pi_{k,h}^{\mathrm{int}}
\triangleq
\frac{
P_k^{[\mathcal{M}_h,\mathrm{int}]}
}{
P_k
} \; .
\end{equation*}
The corresponding net mode-in-variance \ac{pf} is defined as
\begin{equation}
\label{eq:net_stochastic_pf}
\pi_{k,h}^{\mathrm{net}}
\triangleq
\pi_{k,h}^{\mathrm{dir}}
+
\pi_{k,h}^{\mathrm{int}} \; .
\end{equation}
\end{definition}

The net mode-in-variance \ac{pf} 
$\pi_{k,h}^{\mathrm{net}}$
represents the overall contribution of mode $\mathcal{M}_h$ to the stationary variance after accounting for its interactions with other modes.  
From Proposition~\ref{prop:2},
$\pi_{k,h}^{\mathrm{dir}}\geq 0$, whereas 
$\pi_{k,h}^{\mathrm{int}}$
and $\pi_{k,h}^{\mathrm{net}}$
can be positive or negative. For $P_k>0$, the net mode-in-variance \acp{pf} sum to unity by construction, i.e., 
$\sum_h \pi_{k,h}^{\mathrm{net}}
= 1$.

\subsection{Projection to Algebraic Variables}
\label{sec:algebraic_spf}

The proposed mode-in-variance \ac{pf} framework is extended to algebraic variables by accounting for their dependence on the states and stochastic inputs. From the linearized model \eqref{eq:linear_sdae}, the variations of algebraic variables satisfy
\begin{equation}
\label{eq:algebraic_mapping}
\Delta\bfg y
= \bfb G\bfg z \; ,
\end{equation}
where 
\begin{equation}
\label{eq:G_algebraic}
\bfb G
=-\jac{g}{y}^{-1}
\begin{bmatrix}
\jac{g}{x} & \jac{g}{\eta}
\end{bmatrix}
\end{equation}
is the corresponding algebraic mapping matrix. Hence, the stationary covariance matrix of $\Delta\bfg y$ is \cite{adeen2023}
\begin{equation}
\label{eq:algebraic_covariance}
\bfb C
=
\bfb G\bfb P\bfb G^T \; .
\end{equation}
Substituting the decomposition of $\bfb P$ in
\eqref{eq:cov_sum_pairs} into \eqref{eq:algebraic_covariance} gives
\begin{equation}
\label{eq:algebraic_modal_pair}
\bfb C
=
\sum_{i=1}^{\nu}\sum_{j=1}^{\nu}
\Contra_{ij} \, ,
\qquad
\Contra_{ij}
\triangleq
\bfb G\Contr_{ij}\bfb G^T \; .
\end{equation}
where $\Contra_{ij}$ represents the contribution of the eigenvalue pair $(i,j)$ to the stationary covariance of the algebraic variables.

For a $\Delta y_k$ with stationary variance
\begin{equation}
\label{eq:algebraic_variance}
C_k
\triangleq
\bfb C(k,k)
=
\operatorname{Var}(\Delta y_k) \; ,
\end{equation}
the contribution of pair $(i,j)$ is obtained from
\eqref{eq:algebraic_modal_pair} as
%
\begin{equation}
\label{eq:algebraic_modal_pair_variance}
\contrael_k^{(i,j)}
\triangleq
\bfg \epsilon_k^T\Contra_{ij}\bfg \epsilon_k \; ,
\end{equation}
where $\bfg \epsilon_k$ is the $k$th canonical basis vector in $\mathbb{R}^{\ny}$.
Hence,
\begin{equation}
\label{eq:algebraic_variance_modal_decomposition}
C_k
=
\sum_{i=1}^{\nu}\sum_{j=1}^{\nu}
\contrael_k^{(i,j)} \; .
\end{equation}

The eigenvalue pair contributions to $C_k$ are grouped following the
construction in Section~\ref{sec:spf}. The direct and total
interaction contributions associated with mode $\mathcal{M}_h$ are
\begin{equation}
\label{eq:within_mode_variance_y}
C_k^{[\mathcal{M}_h,\mathrm{dir}]}
\triangleq
\sum_{i\in\mathcal{M}_h}
\sum_{j\in\mathcal{M}_h}
\contrael_k^{(i,j)} \; ,
\end{equation}
and
\begin{equation}
\label{eq:mode_interaction_variance_y}
C_k^{[\mathcal{M}_h,\mathrm{int}]}
\triangleq
\frac{1}{2}\sum_{\ell\neq h}
\sum_{i\in\mathcal{M}_h}
\sum_{j\in\mathcal{M}_\ell}
\left(
\contrael_k^{(i,j)}
+
\contrael_k^{(j,i)}
\right)\; .
\end{equation}

Thus, the direct and interaction mode-in-variance \acp{pf} for $\Delta y_k$ are defined as
\begin{equation*}
 \pi_{k,h}^{y,\mathrm{dir}}
\triangleq
\frac{
C_k^{[\mathcal{M}_h,\mathrm{dir}]}
}{
C_k
}\ , \qquad
\pi_{k,h}^{y,\mathrm{int}}
\triangleq
\frac{
C_k^{[\mathcal{M}_h,\mathrm{int}]}
}{
C_k
}\ .
\end{equation*}
The corresponding net mode-in-variance PF is defined as:
\begin{equation}
\pi_{k,h}^{y,\mathrm{net}}
\triangleq
\pi_{k,h}^{y,\mathrm{dir}}
+
\pi_{k,h}^{y,\mathrm{int}} \; .
\end{equation}
For $C_k>0$, the net mode-in-variance \acp{pf} sum to unity by construction, i.e.,
$\sum_h \pi_{k,h}^{y,\mathrm{net}} = 1$.

\section{Case Studies}
\label{sec:case_studies}

This section illustrates the proposed mode-in-variance \ac{pf} framework through numerical case studies based on the Kundur two-area and IEEE 39-bus benchmark systems. All simulations are performed using Dome \cite{milano2022power}.

\subsection{Kundur Two-Area System}

\begin{figure}[ht!]
    \centering    \includegraphics[width=0.9\linewidth]{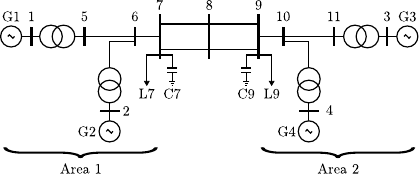}
    \caption{Kundur two-area system: single-line diagram.}
    \label{fig1:kundur}
\end{figure}

This section considers the two-area four-machine system shown in Fig.~\ref{fig1:kundur}. The system consists of eleven buses and four \acp{sm} connected through a relatively weak tie line. Each \ac{sm} is equipped with an \ac{avr} and a \ac{tg}. The system feeds two loads, modeled as constant power demands subject to independent stochastic variations represented by OU processes. Considering active and reactive power variations per load gives $\neta=4$. This leads to a special case of \eqref{eq:oup} with
\[
\bfb A_\eta
\!=\!
\mathrm{diag}(\alpha_{p_1},\alpha_{q_1},\alpha_{p_2},\alpha_{q_2}) ,
\bfb B_\eta
\!=\!
\mathrm{diag}(\beta_{p_1},\beta_{q_1},\beta_{p_2},\beta_{q_2}) \, .
\]

The coefficients $\alpha_p$ and $\alpha_q$ determine the autocorrelation decay rates of the active and reactive power variations, respectively, while
$\beta_p=\sigma_p\sqrt{2\alpha_p}$ and
$\beta_q=\sigma_q\sqrt{2\alpha_q}$, where $\sigma_p$ and $\sigma_q$ are the corresponding stationary standard deviations.  We set $\alpha_p=0.1~\mathrm{s}^{-1}$ and $\sigma_p=0.08p_{L0}$ for the active power variations, and $\alpha_q=0.01~\mathrm{s}^{-1}$ and $\sigma_q=0.01q_{L0}$ for the reactive power variations.

\begin{figure}[ht!]
    \centering
    \subfloat[Mode~27]{   \includegraphics[width=0.47\linewidth]{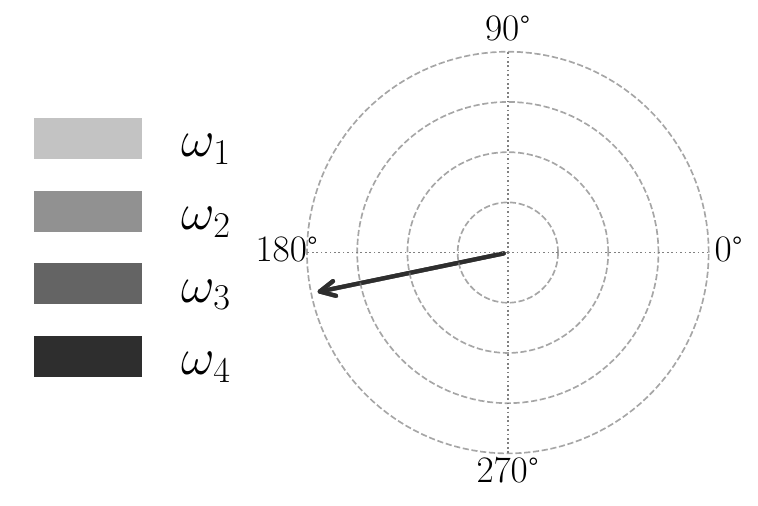}
        \label{fig:mode1}
    }
    \subfloat[Mode~17]{      \includegraphics[width=0.47\linewidth]{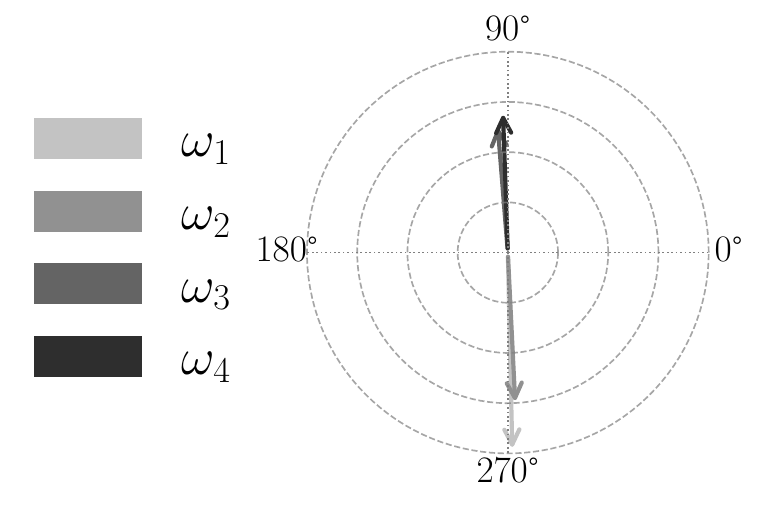}
    \label{fig:mode2}
    }
    \caption{Kundur system: Mode shapes (a) common mode and (b) interarea.}  \label{fig:kundur_modes}
\end{figure}

The decomposition in \eqref{eq:variance_mode_decomposition} is first validated by comparing the variances obtained by summing the direct and interaction contributions of all modes with the reference values obtained by solving \eqref{eq:lyapunov} using SLICOT \cite{Benner1999SLICOT}. In addition, the variances are also validated against the reference values estimated from a 10,000~s nonlinear \ac{tds}. For each comparison, the relative error is calculated as the $2$-norm of the difference between the variance vectors, normalized by the $2$-norm of the corresponding reference vector. The error relative to the Lyapunov solution is as expected negligible, at $2.6\times10^{-12}\%$, while the error relative to the \ac{tds} is $3.696\%$,
confirming the numerical accuracy of the decomposition and 
indicating good agreement with the simulated responses.

The proposed mode-in-variance \acp{pf} are used to identify the dynamic modes contributing to the stationary variability of the rotor speed of \ac{sm}~1. 
The stationary standard deviation of $\omega_1$ is $\sigma_{\omega_1}= 0.4695 \times 10^{-3} ~\mathrm{pu}$. 
The results are reported in Table~\ref{tab:stochastic_pf_state_4}, which lists the five modes with the largest absolute net contributions.

\begin{table}[ht!]
\centering
\caption{Kundur system: mode-in-variance \acp{pf} for $\omega_1$.}
\label{tab:stochastic_pf_state_4}
\resizebox{\columnwidth}{!}{%
\begin{tabular}{c c c c c c c}
\hline
Mode & $\lambda$ & $f_n$ (Hz) & $\zeta$ & $\pi^{\mathrm{dir}}$ & $\pi^{\mathrm{int}}$ & $\pi^{\mathrm{net}}$ \\
\hline
27 & $-0.322 \pm j0.347$ & 0.075 & 0.680 & 0.9086 & 0.0249 & 0.9335 \\
17 & $-0.117 \pm j3.126$ & 0.498 & 0.037 & 0.0422 & -0.0025 & 0.0397 \\
23 & $-1.406$ & 0.224 & 1.000 & 0.0082 & 0.0122 & 0.0204 \\
24 & $-0.850$ & 0.135 & 1.000 & 0.0006 & 0.0043 & 0.0049 \\
28 & $-0.363 \pm j0.298$ & 0.075 & 0.773 & 0.0001 & -0.0048 & -0.0047 \\
\hline
\end{tabular}}
\end{table}

Mode~27, represented by the eigenvalue pair $-0.322 \pm j0.347$, has the largest direct and net mode-in-variance \acp{pf} in $\omega_1$. The shapes of the two oscillatory modes with positive net contributions are shown in Fig.~\ref{fig:kundur_modes}. With a natural frequency of $0.075$~Hz, Mode~27 exhibits a coherent shape across the rotor speeds of all machines, corresponding to the common mode of the system~\cite{chow2016,2019_kamwa,bouterakos2025}. In contrast, the second-largest contributor, Mode~17, corresponds to the interarea mode. With a $\pi^{\mathrm{net}}=0.9335$, compared with $0.0397$ for the interarea mode, Mode~27 dominates the stationary variability of $\omega_1$. 
Notably, Mode~27 has a high damping ratio $\zeta = 0.680$ and thus, although not critical from the perspective of conventional small-signal screening, is highly relevant under stochastic excitation. Moreover, the small value of
$\pi^{\mathrm{int}}=0.0497$ indicates that interactions with the remaining modes slightly influence the variability of $\omega_1$.

\begin{table}[ht!]
\centering
\caption{Kundur system: mode-in-variance \acp{pf} for $\omega_{\mathrm{COI}}$.}
\label{tab:stochastic_pf_algebraic_88}
\resizebox{\columnwidth}{!}{%
\begin{tabular}{c c c c c c c}
\hline
Mode & $\lambda$ & $f_n$ (Hz) & $\zeta$ & $\pi^{y,\mathrm{dir}}$ & $\pi^{y,\mathrm{int}}$ & $\pi^{y,\mathrm{net}}$ \\
\hline
27 & $-0.322 \pm j0.347$ & 0.075 & 0.680 & 0.9542 & 0.0191 & 0.9733 \\
23 & $-1.406$ & 0.224 & 1.000 & 0.0086 & 0.0153 & 0.0238 \\
24 & $-0.850$ & 0.135 & 1.000 & 0.0008 & 0.0056 & 0.0064 \\
28 & $-0.363 \pm j0.298$ & 0.075 & 0.773 & 0.0000 & -0.0040 & -0.0040 \\
17 & $-0.117 \pm j3.126$ & 0.498 & 0.037 & 0.0006 & -0.0001 & 0.0005 \\
\hline
\end{tabular}}
\end{table}

The proposed framework is further illustrated by considering the \ac{coi} frequency of the system, which is an algebraic variable.
Its stationary standard deviation is $\sigma_{\omega_{\mathrm{COI}}} = 0.4580 \times 10^{-3}~\mathrm{pu}$, 
and the corresponding mode-in-variance \acp{pf} are reported in Table~\ref{tab:stochastic_pf_algebraic_88}. As can be seen, the common mode again provides the largest direct and net contributions, with $\pi^{y,\mathrm{net}}=0.9733$,
indicating that it is the dominant contributor to stochastic frequency variability. 

The information provided by the proposed mode-in-variance \acp{pf} is next contrasted with that of classical mode-in-state \acp{pf}.
In particular, Table~\ref{tab:classical_pf_state_omega_syn01} reports the classical \acp{pf} for the rotor speed of \ac{sm}~1. The column "dom." indicates the state in which each mode exhibits the largest participation. The local mode at $f_n=1.054$~Hz and interarea mode exhibit the largest classical \acp{pf}, while the common mode at $f_n=0.075$~Hz ranks third.  This difference results from the distinct quantities characterized by the two formulations.  The larger classical \acp{pf} of the local and interarea modes indicate their greater contributions to the response of $\omega_1$ following an initial perturbation of that variable. In contrast, the mode-in-variance \acp{pf} identify the common mode as the dominant contributor to the stationary variance of $\omega_1$ under continuous stochastic excitation.

\begin{table}[ht!]
\centering
\caption{Kundur system: Classical 
\acp{pf} for $\omega_{1}$.}
\label{tab:classical_pf_state_omega_syn01}
\resizebox{\columnwidth}{!}{%
\begin{tabular}{c l c c c c}
\hline
Mode & dom. & $\lambda$ & $f_n$ (Hz) & $\zeta$ & \ac{pf} \\
\hline
18 & $\omega_2$  & $-0.405 \pm j6.608$ & 1.054 & 0.061 & 0.2951 \\
17 & $\omega_1$  & $-0.117 \pm j3.127$ & 0.498 & 0.037 & 0.2836 \\
27 & $\omega_3$ & $-0.322 \pm j0.347$ & 0.075 & 0.68 & 0.1906 \\
23 & $\omega_1$ & $-1.406$ & 0.224 & 1 & 0.1640 \\
16 & $\omega_4$ & $-0.404 \pm j6.753$ & 1.077 & 0.06 & 0.0895 \\
\hline
\end{tabular}}
\end{table}
\color{black}

\subsection{Modified Kundur Two-Area System}

This section considers a modified version of the Kundur two-area system, in which the \acp{sm} in Area~1 are replaced by \ac{gfl} converters modeled with 
a synchronous-reference-frame Phase-Locked Loop (SRF-PLL), 
inner PI current control loops, and constant DC voltage \cite{ignacio_2025}. The
converters are equipped with 
frequency controllers consisting of a droop gain $R=0.05$ and a first-order filter with time constant $T_f = 2~s$. The stochastic load model and parameters are kept the same as in the previous section.

\begin{figure}[ht!]
  \centering    \includegraphics[width=0.9\linewidth]{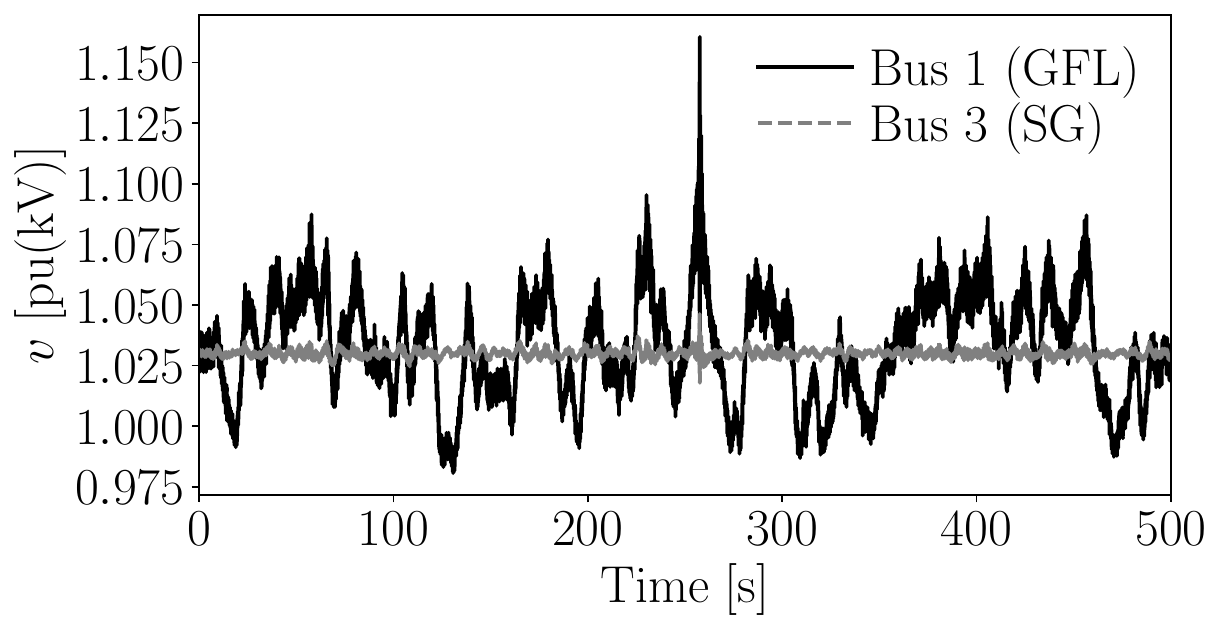}
  
  \vspace{-3mm}
  \caption{Modified Kundur system: Voltage responses at Buses 1 and 3.}   \label{fig:voltage_comparison}
\end{figure}

We examine the stochastic voltage response at two representative buses of the modified system: Bus~1, i.e., the terminal bus of a \ac{gfl} converter, and Bus~3, i.e., the terminal bus of a \ac{sm}. 
Figure~\ref{fig:voltage_comparison} shows a realization of the two voltage responses obtained from a nonlinear \ac{tds} of \eqref{eq:sdae}.  The voltage at the \ac{gfl} terminal exhibits substantially larger variability than that at the \ac{sm} terminal, with stationary standard deviations of $\sigma_{V_1}=0.02~\mathrm{pu}$ and $\sigma_{V_3}=0.0014~\mathrm{pu}$, respectively.

The mode-in-variance \acp{pf} for $V_1$ are reported in Table~\ref{tab:stochastic_pf_algebraic_V1}. Mode~17, whose classical \acp{pf} indicate dominant participation to the \ac{gfl} frequency control dynamics, has the largest net contribution, with $\pi^{y,\mathrm{net}}=0.9233$.
 Its direct contribution is larger, with $\pi^{y,\mathrm{dir}}=1.1719$, but is partially offset by the negative interaction contribution, $\pi^{y,\mathrm{int}}=-0.2486$.  The second-largest positive contribution is provided by Mode~25.

\begin{table}[htb!]
\centering
\caption{Modified Kundur system: mode-in-variance \acp{pf} for $V_{1}$.}
\label{tab:stochastic_pf_algebraic_V1}
\resizebox{\columnwidth}{!}{%
\begin{tabular}{c c c c c c c}
\toprule
Mode & $\lambda$ & $f_n$ (Hz) & $\zeta$ & $\pi^{y,\mathrm{dir}}$ & $\pi^{y,\mathrm{int}}$ & $\pi^{y,\mathrm{net}}$ \\
\midrule
17 & $-0.143$ & 0.023 & 1.000 & 1.1719 & -0.2486 & 0.9233 \\
25 & $-0.100$ & 0.016 & 1.000 & 0.0252 & 0.1222 & 0.1474 \\
24 & $-0.100$ & 0.016 & 1.000 & 0.0999 & -0.1361 & -0.0362 \\
14 & $-0.529 \pm j0.444$ & 0.110 & 0.766 & 0.0350 & -0.0683 & -0.0333 \\
1 & $-59.311$ & 9.440 & 1.000 & 0.0204 & 0.0019 & 0.0223 \\
\bottomrule
\end{tabular}}
\end{table}

\begin{table}[ht!]
\centering
\caption{Modified Kundur system: mode-in-variance \acp{pf} for $V_{3}$.}
\label{tab:stochastic_pf_algebraic_V3}
\resizebox{\columnwidth}{!}{%
\begin{tabular}{c c c c c c c}
\toprule
Mode & $\lambda$ & $f_n$ (Hz) & $\zeta$ & $\pi^{y,\mathrm{dir}}$ & $\pi^{y,\mathrm{int}}$ & $\pi^{y,\mathrm{net}}$ \\
\midrule
13 & $-0.364 \pm j1.009$ & 0.171 & 0.339 & 2.4606 & -1.7440 & 0.7167 \\
1 & $-59.311$ & 9.440 & 1.000 & 0.2099 & -0.0262 & 0.1837 \\
14 & $-0.529 \pm j0.444$ & 0.110 & 0.766 & 3.5554 & -3.4414 & 0.1140 \\
18 & $-0.353 \pm j0.292$ & 0.073 & 0.771 & 0.1079 & -0.1971 & -0.0892 \\
17 & $-0.143$ & 0.023 & 1.000 & 0.1086 & -0.0757 & 0.0329 \\
\bottomrule
\end{tabular}}
\end{table}

The mode-in-variance \acp{pf} for $V_3$ are reported in Table~\ref{tab:stochastic_pf_algebraic_V3}. Mode~13, corresponding to the system's common mode, exhibits the largest net contribution, with $\pi^{y,\mathrm{net}}=0.7167$. In contrast, Mode~14, associated with the \ac{sm} voltage control dynamics, has the largest direct contribution, $\pi^{y,\mathrm{dir}}=3.5554$, but its strong negative interaction contribution reduces its net \ac{pf} to $\pi^{y,\mathrm{net}}=0.1140$. Thus, a large direct contribution does not necessarily translate into a large overall contribution, highlighting the importance of accounting for modal interactions when attributing stationary voltage variability.

For comparison, the classical mode-in-algebraic variable \acp{pf} for $V_{1}$ are calculated using the formulation in \cite{tzounas2019modal} and reported in Table~\ref{tab:classical_pf_algebraic_v_bus_01}. 
The modes with the largest mode-in-algebraic \acp{pf} 
exhibit their dominant state participation in the internal transient voltage states of the \acp{sm}, namely $e'_{q,4}$, $e'_{d,3}$, the d-axis PI state of the \ac{gfl} $x_{d,{\mathrm{GFL}2}}$, the SRF-PLL state $\theta_{\mathrm{PLL1}}$, and the \ac{avr} state $v_{r,3}$.
In contrast, the mode-in-variance \acp{pf} identify Mode~17, associated with the \ac{gfl} frequency control dynamics, as the dominant contributor to the voltage variance at Bus~1. 
These results further confirm that classical \acp{pf} should not be used to assess modal contributions to variability under continuous stochastic excitation.

\begin{table}[ht!]
\centering
\caption{Modified Kundur system: Classical \acp{pf} for $V_{1}$.}
\label{tab:classical_pf_algebraic_v_bus_01}
\resizebox{\columnwidth}{!}{%
\begin{tabular}{c l c c c c}
\hline
Mode  & dom. &  $\lambda$ & $f_n$ (Hz) & $\zeta$ & \ac{pf} \\
\hline
18 & $e'_{q,4}$ & $-0.353\pm j0.292$ & 0.073 & 0.771 & 0.0657 \\
16 & $e'_{d,3}$ & $-4.840$ & 0.77 & 1 & 0.0425 \\
33 & $x_{d,{\mathrm{GFL}2}}$ & $-0.500$ & 0.08 & 1 & 0.0362 \\
1 & $\theta_{\mathrm{PLL},1}$ & $-59.311$ & 9.44 & 1 & 0.0338 \\
14 & $v_{r,3}$ & $-0.529\pm j0.444$ & 0.11 & 0.766 & 0.0261 \\
\hline
\end{tabular}}
\end{table}

\begin{figure}[ht!]
  \centering  \includegraphics[width=0.9\linewidth]{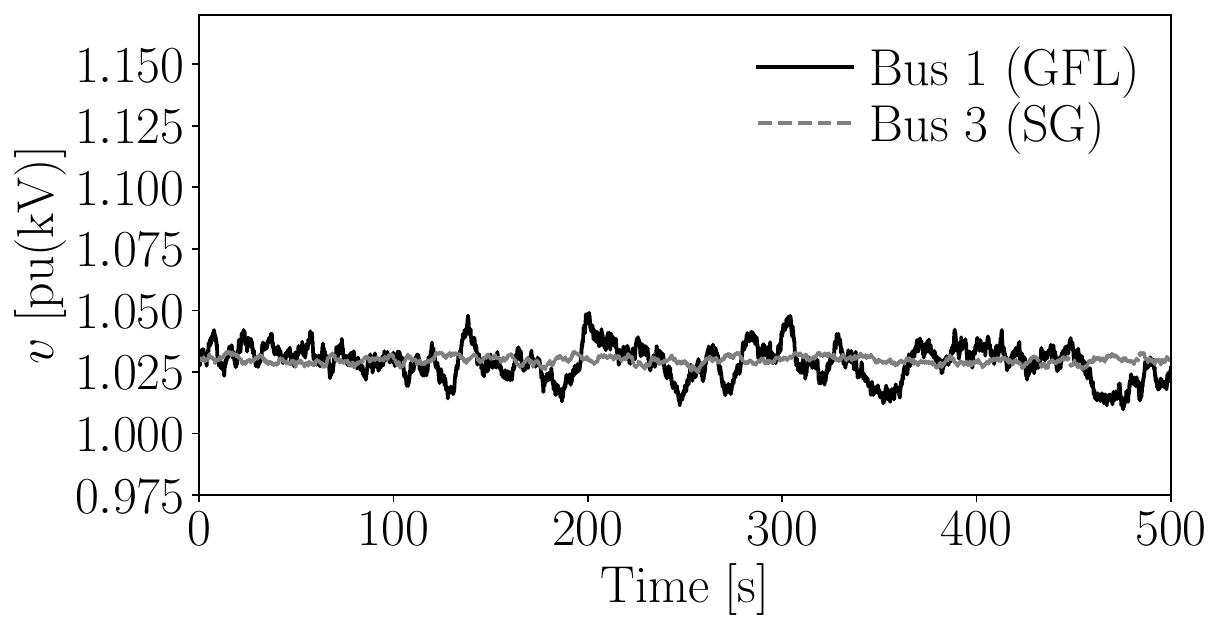}
  \caption{Modified Kundur system: GFL without primary frequency control: voltage at the terminal buses of the GFL at Bus~1 and \ac{sm} at Bus~3.}  \label{fig:voltage_comparison2}
\end{figure}

To further examine the relationship between frequency control and voltage variability, a second configuration is considered in which the frequency control of the \acp{gfl} is disabled. The resulting terminal bus voltage responses of the \ac{gfl} at Bus~1 and \ac{sm} at Bus~3 are shown in Fig.~\ref{fig:voltage_comparison2}.
In this scenario, the stationary 
standard deviation
of $V_1$ decreases to 
$0.007$~${\rm pu}$,
corresponding to $65\%$
reduction relative to the previous configuration. 
The corresponding mode-in-variance \acp{pf} are reported in Table~\ref{tab:stochastic_pf_algebraic_23}. Modes~23 and~24, associated with the stochastic active load processes, now provide the largest $\pi^{y,\mathrm{net}}$, equal to $0.6844$ and $0.3021$, respectively.  Notably, both modes have damping ratios $\zeta=1$, further showing that non-oscillatory modes, often receiving less attention in conventional small-signal stability assessments, can dominate stationary voltage variability under stochastic excitation.

\begin{table}[ht!]
\centering
\caption{Modified Kundur system: mode-in-variance \acp{pf} for $V_{1}$.}
\label{tab:stochastic_pf_algebraic_23}
\resizebox{\columnwidth}{!}{%
\begin{tabular}{c c c c c c c}
\toprule
Mode & $\lambda$ & $f_n$ (Hz) & $\zeta$ & $\pi^{y,\mathrm{dir}}$ & $\pi^{y,\mathrm{int}}$ & $\pi^{y,\mathrm{net}}$ \\
\midrule
23 & $-0.100$ & 0.016 & 1.000 & 0.7805 & -0.0961 & 0.6844 \\
24 & $-0.100$ & 0.016 & 1.000 & 0.2409 & 0.0612 & 0.3021 \\
14 & $-0.523 \pm j0.437$ & 0.109 & 0.767 & 0.0333 & -0.0181 & 0.0153 \\
19 & $-0.491$ & 0.078 & 1.000 & 0.0002 & -0.0059 & -0.0057 \\
11 & $-4.594$ & 0.731 & 1.000 & 0.0014 & 0.0038 & 0.0052 \\
\bottomrule
\end{tabular}}
\end{table}

We note that the stochastic load model and parameters remain unchanged between the two configurations considered above, despite the substantial reduction in voltage variability.  This illustrates that the magnitude of stochastic variability cannot be inferred solely from the characteristics of the stochastic inputs, since it strongly depends on how the system dynamics shape their propagation to the variable of interest.  In this regard, the proposed \acp{pf} can be used to identify control dynamics that can be targeted to reduce stochastic variability in electrical quantities of interest. This is further discussed in the next section.

\subsection{IEEE 39-Bus System}

\begin{figure}[ht!]
  \centering
\includegraphics[width=0.85\linewidth]{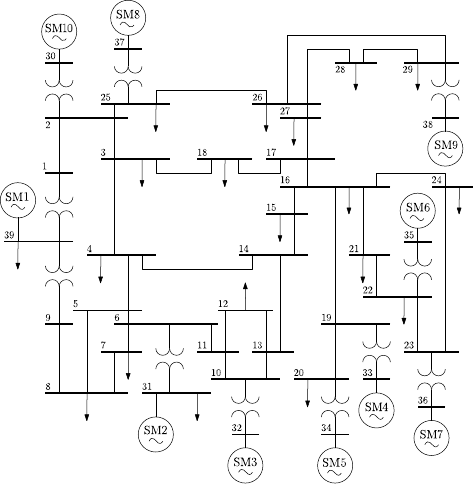}
  \caption{IEEE 39-bus system: single-line diagram.}
  \label{fig:39b_topo}
\end{figure}

This section considers the standard IEEE 39-bus benchmark system shown in Fig.~\ref{fig:39b_topo}.
It consists of ten \acp{sm},  
all represented with fourth-order models and equipped with voltage regulators, \acp{tg}, and PSSs.  For the purposes of this paper, all 19 loads in the system are modeled by OU processes, with the same parameters for all loads. For the active power variations, $\alpha_p=1~\mathrm{s}^{-1}$ and the standard deviation is set to $\sigma_p=0.05p_{L0}$. For the reactive power variations, $\alpha_q=1~\mathrm{s}^{-1}$ and $\sigma_q=0.01q_{L0}$.

\begin{figure}[ht!]
  \centering  \includegraphics[width=0.9\linewidth]{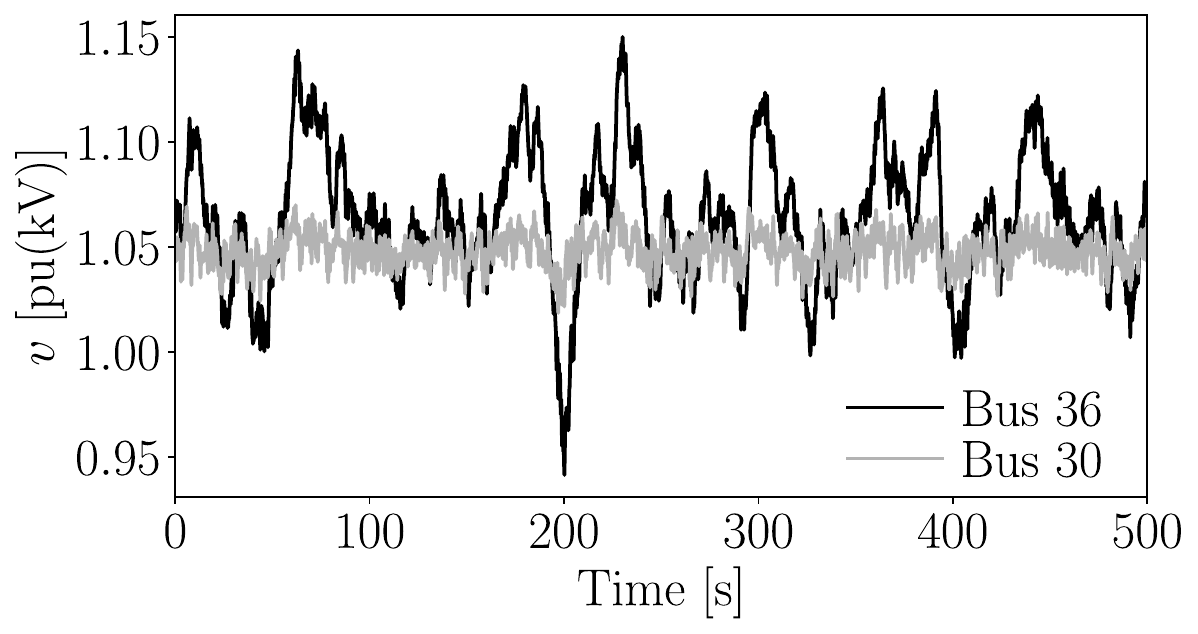}

  \vspace{-3mm}
  \caption{39-bus system: 
  Highest and lowest bus voltage variances.}  \label{fig:voltage_comparison_39bus}
\end{figure}

The stochastic voltage variability is evaluated at the terminal buses of all ten \acp{sm}. Among these buses, $V_{36}$ has the highest stationary standard deviations, $\sigma_{V_{36}}= 0.0258 ~\mathrm{pu}$, whereas $V_{30}$ has the lowest, $\sigma_{V_{30}}= 0.0072~\mathrm{pu}$. The voltage responses at these two buses for one stochastic realization are shown in Fig.~\ref{fig:voltage_comparison_39bus}.

\begin{figure}[ht!]
    \centering
    \subfloat[Mode~54]{       \includegraphics[width=0.45\linewidth]{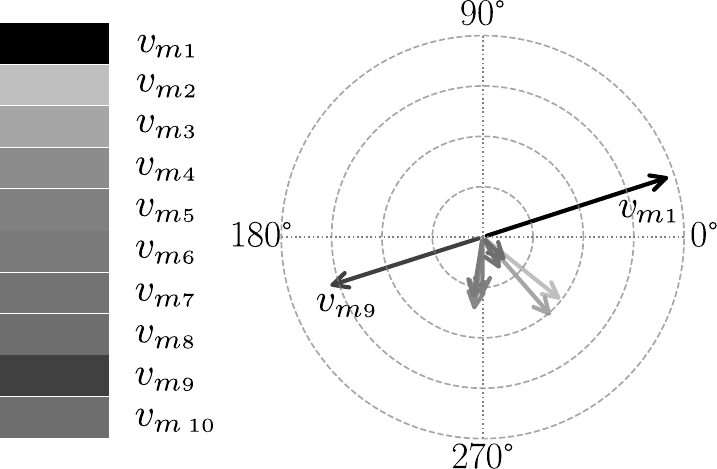}
        \label{fig:mode54}
    }
    \subfloat[Mode~55]{        \includegraphics[width=0.45\linewidth]{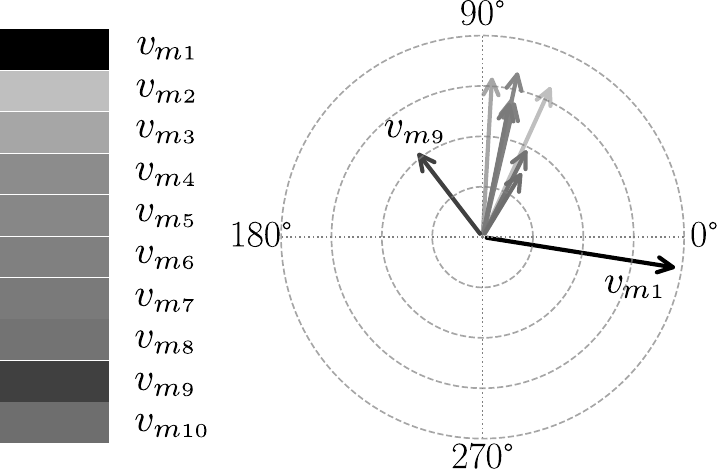}
        \label{fig:mode55}
    }
    \caption{39-bus system: Mode shapes for (a) Mode~54 and (b) Mode~55.
    }
    \label{fig:39b_modes}
\end{figure}

Modes~54 and~55, with high damping ratios of $\zeta=0.964$ and $0.996$, respectively, consistently appear among those with the largest net mode-in-variance \acp{pf} for the voltages across the \ac{sm} terminal buses. Fig.~\ref{fig:39b_modes} shows the shapes for Mode~54 and 55, using the eigenvectors related to the voltage states measured by the corresponding \ac{sm} \acp{avr} ($v_{m}$). The mode shapes exhibit a predominantly anti-phase response between \acp{sm}~1 and~9.
The mode-in-variance \acp{pf} for $V_{30}$ and $V_{36}$ are reported in Tables~\ref{tab:stochastic_pf_algebraic_V30} and \ref{tab:stochastic_pf_algebraic_V36}, respectively. Mode~54 exhibits the largest net contribution at both buses. 

\begin{table}[ht!]
\centering
\caption{39-bus system: mode-in-variance \acp{pf} for $V_{30}$.}
\label{tab:stochastic_pf_algebraic_V30}
\resizebox{\columnwidth}{!}{%
\begin{tabular}{c c c c c c c}
\toprule
Mode & $\lambda$ & $f_n$ (Hz) & $\zeta$ & $\pi^{y,\mathrm{dir}}$ & $\pi^{y,\mathrm{int}}$ & $\pi^{y,\mathrm{net}}$ \\
\midrule
54 & $-1.311 \pm j0.363$ & 0.216 & 0.964 & 39.1177 & -34.1835 & 4.9342 \\
55 & $-1.550 \pm j0.145$ & 0.248 & 0.996 & 30.1559 & -34.3772 & -4.2213 \\
47 & $-1.359 \pm j1.205$ & 0.289 & 0.748 & 1.3636 & -0.3528 & 1.0108 \\
60 & $-1.384$ & 0.220 & 1.000 & 1.1032 & -1.9298 & -0.8266 \\
49 & $-1.362 \pm j0.697$ & 0.243 & 0.890 & 1.1595 & -1.8219 & -0.6624 \\
\bottomrule
\end{tabular}}
\end{table}

\begin{table}[ht!]
\centering
\caption{39-bus system: mode-in-variance \acp{pf} for $V_{36}$.}
\label{tab:stochastic_pf_algebraic_V36}
\resizebox{\columnwidth}{!}{%
\begin{tabular}{c c c c c c c}
\toprule
Mode & $\lambda$ & $f_n$ (Hz) & $\zeta$ & $\pi^{y,\mathrm{dir}}$ & $\pi^{y,\mathrm{int}}$ & $\pi^{y,\mathrm{net}}$ \\
\midrule
54 & $-1.311 \pm j0.363$ & 0.216 & 0.964 & 4.6747 & -3.7215 & 0.9532 \\
55 & $-1.550 \pm j0.145$ & 0.248 & 0.996 & 6.2868 & -7.206 & -0.9192 \\
79 & $-0.097$ & 0.015 & 1.000 & 2.0632 & -1.2690 & 0.7942 \\
49 & $-1.362 \pm j0.697$ & 0.243 & 0.890 & 4.2394 & -4.8933 & -0.6538 \\
47 & $-1.359 \pm j1.205$ & 0.289 & 0.748 & 2.8841 & -2.5197 & 0.3644 \\
\bottomrule
\end{tabular}}
\end{table}

The sign of the net mode-in-variance \ac{pf} provides additional insight into the role of modal interactions. 
In particular, at Bus~38, Mode~49 has the largest positive net contribution, whereas Mode~54 has a large direct contribution of $42.5011$, but its negative interaction contribution of $-46.6281$ results in $\pi^{y,\mathrm{net}}=-4.1270$ (Table~\ref{tab:stochastic_pf_algebraic_V38}). Notably, Bus~38 is the only \ac{sm} terminal bus where the interactions of Mode~54 with the remaining modes outweigh its direct contribution, resulting in a net reduction in the total voltage variance. 

\begin{table}[ht!]
\centering
\caption{39-bus system: mode-in-variance \acp{pf} for $V_{38}$.}
\label{tab:stochastic_pf_algebraic_V38}
\resizebox{\columnwidth}{!}{%
\begin{tabular}{c c c c c c c}
\toprule
Mode & $\lambda$ & $f_n$ (Hz) & $\zeta$ & $\pi^{y,\mathrm{dir}}$ & $\pi^{y,\mathrm{int}}$ & $\pi^{y,\mathrm{net}}$ \\
\midrule
49 & $-1.362 \pm j0.697$ & 0.243 & 0.890 & 131.9305 & -123.2937 & 8.6368 \\
54 & $-1.311 \pm j0.363$ & 0.216 & 0.964 & 42.5011 & -46.6281 & -4.1270 \\
55 & $-1.550 \pm j0.145$ & 0.248 & 0.996 & 2.5377 & -3.8279 & -1.2902 \\
63 & $-1.070$ & 0.170 & 1.000 & 2.4507 & -3.6901 & -1.2394 \\
60 & $-1.384$ & 0.220 & 1.000 & 1.7542 & -0.7145 & 1.0397 \\
\bottomrule
\end{tabular}}
\end{table}

The classical mode-in-algebraic variable \acp{pf} are also evaluated for $V_{36}$, with the results reported in Table~\ref{tab:classical_pf_algebraic_v_bus_36}. The largest \acp{pf} are associated with generator internal-voltage states, including $e'_{\rm d}$ of \acp{sm}~4 and 7 and $e'_{\rm q}$ of \acp{sm}~6 and 7, as well as the $\rm q$-axis
current $i_{q,6}$ of \ac{sm}~6. Notably, neither Mode~54 nor Mode~55 appears among the modes with the largest \acp{pf}. 

\begin{table}[ht!]
\centering
\caption{39-bus system: Classical \acp{pf} for $V_{36}$.}
\label{tab:classical_pf_algebraic_v_bus_36}
\resizebox{\columnwidth}{!}{%
\begin{tabular}{c l c c c c}
\hline
Mode & dom. &  $\lambda$ & $f_n$ (Hz) & $\zeta$ & \ac{pf} \\
\hline
93 & $i_{q,6}$ & $-0.916$ & 0.146 & 1 & 0.3042 \\
92 & $e'_{d,4}$ & $-0.959$ & 0.153 & 1 & 0.2861 \\
63 & $e'_{d,9}$ & $-1.070$ & 0.170 & 1 & 0.0636 \\
86 & $e'_{q,6}$  & $-0.486\pm j0.628$ & 0.126 & 0.612 & 0.0107 \\
70 & $e'_{d,7}$ & $-0.806\pm j0.819$ & 0.183 & 0.701 & 0.0107 \\
\hline
\end{tabular}}
\end{table}

\begin{figure}[ht!]
  \centering  \includegraphics[width=0.9\linewidth]{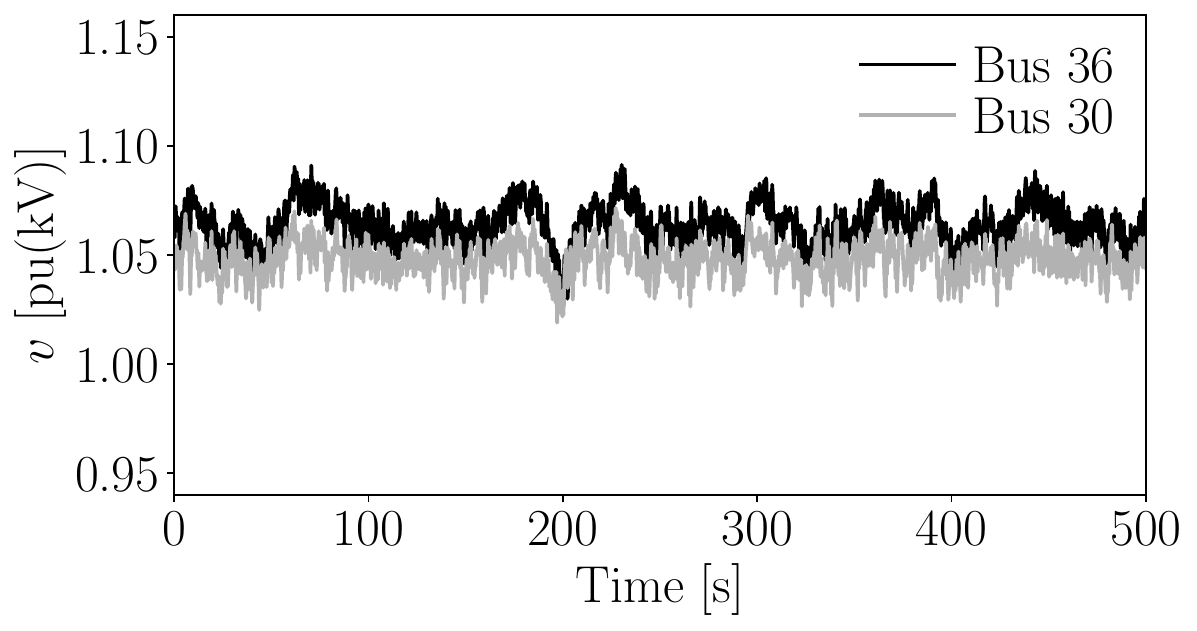}

  \vspace{-3mm}
  \caption{39-bus system: voltage magnitude variations after the adjustment in the PSS gain of the \ac{sm} at Bus 36.}
\label{fig:voltage_comparison_39busmod}
\end{figure}

For Bus~36, Table~\ref{tab:stochastic_pf_algebraic_V36} identifies, in addition to Modes~54 and 55
discussed above, Mode~79 as a significant positive net contributor to voltage variance, with $\pi^{y,\mathrm{net}}=0.7942$. Mode~79 is a local mode associated with the PSS dynamics of the \ac{sm} connected to Bus~36.  Based on this, we finally investigate whether a retuning of the 
PSS can be beneficial in reducing the stationary voltage variability at Bus~36.  As shown in Fig.~\ref{fig:voltage_comparison_39busmod}, reducing the PSS gain from 7.5 to 2 decreases the stationary standard deviation from $\sigma_{V_{36}}=0.0258~\mathrm{pu}$ to $\sigma_{V_{36}}=0.0083~\mathrm{pu}$, corresponding to a 67.83\% reduction in the local voltage variability at bus 36.

\section{Conclusions}
\label{sec:conclu}

The variability of a power system quantity cannot be attributed solely to the characteristics of stochastic inputs, since the system dynamics shape how disturbances propagate and affect its response. To investigate this relationship, a mode-in-variance \ac{pf} framework was proposed based on the modal decomposition of the stationary covariance matrix obtained from a linearized \ac{sdae} system model. The proposed formulation quantifies the contribution of dynamic modes to the stationary variance of both state and algebraic variables under continuous stochastic excitation, distinguishing their direct contributions from the effects of modal interactions.

The case studies demonstrated that 
highly damped modes, which may
receive less attention in conventional small-signal stability assessments, can dominate stationary frequency and voltage variability. Comparisons with classical \acp{pf} further showed that the modes contributing most strongly to stationary variance do not necessarily coincide with those exhibiting the largest classical participation factors.  Moreover, the proposed framework identified control dynamics whose modification can result in reductions in voltage variability, demonstrating its potential to guide targeted control interventions.  We will dedicate future work to developing this capability into a systematic control design aimed at reducing the stationary variance of selected power system variables.



\appendix

\subsection{Derivation of \texorpdfstring{\eqref{eq:lyapunov}}{the Lyapunov equation}}
\label{app:lyapderiv}

Since the Wiener increments have zero mean, i.e., $\mathbb{E}[d\bfg w]=\bfg 0$, taking expectations in \eqref{eq:linear_sdae} gives
 \begin{align}
 \label{eq:dEz}
 d\mathbb{E}[\bfg z]
 =
 \bfb A_0\mathbb{E}[\bfg z]\,dt \, ,
 \end{align}
whose solution is \eqref{eq:Ez}.
Consider now the centered variable
\begin{align}
 \label{eq:centered}
 \tilde{\bfg z} = \bfg z-\mathbb{E}[\bfg z] \,  .
 \end{align}

From \eqref{eq:cov_definition}, we have
\begin{equation}
\label{eq:P_defc}
\bfb P(t)
=
\mathbb{E}
\left[
\tilde{\bfg z}(t)
\tilde{\bfg z}^T(t)
\right].
\end{equation}

 Differentiating \eqref{eq:centered} and using \eqref{eq:linear_sdae} and \eqref{eq:dEz} gives:
 \begin{align}
 \label{eq:dzc}
 d\tilde{\bfg z}
 =
 \bfb A_0\tilde{\bfg z}\,dt
 +
 \bfb B_0\,d\bfg w.
 \end{align}

Applying Itô's product rule to $\tilde{\bfg z}\tilde{\bfg z}^T$ yields
\begin{align}
d\left(\tilde{\bfg z}\tilde{\bfg z}^T\right)
={}&
 d\tilde{\bfg z}\,\tilde{\bfg z}^T
 +
 \tilde{\bfg z}\,d\tilde{\bfg z}^T
 +
 d\tilde{\bfg z}\,d\tilde{\bfg z}^T.
 \label{eq:ito_product}
 \end{align}
 Substituting \eqref{eq:dzc} and using 
 $d\bfg w\,d\bfg w^T=\bfb I\,dt$
yields
 \begin{align}
 d\left(\tilde{\bfg z}\tilde{\bfg z}^T\right) ={}&
 \bfb A_0\tilde{\bfg z}\tilde{\bfg z}^T\,dt
 +
 \tilde{\bfg z}\tilde{\bfg z}^T\bfb A_0^T\,dt
 +
 \bfb B_0\bfb B_0^T\,dt
 \nonumber\\
 &+
 \bfb B_0\,d\bfg w\,\tilde{\bfg z}^T +
 \tilde{\bfg z}\,d\bfg w^T\bfb B_0^T .
 \label{eq:ito_cov}
 \end{align}
Taking expectations and 
using \eqref{eq:P_defc} gives
%
\begin{align}
{d\bfb P(t)}/{dt} =
\bfb A_0\bfb P(t) +
\bfb P(t)\bfb A_0^T +
\bfb Q \, .
\label{eq:cov_dynamics}
\end{align}
Since $\bfb A_0$ is Hurwitz, $\bfb P(t)$ converges to its stationary value as $t\rightarrow\infty$, and $d\bfb P(t)/dt\rightarrow\bfb 0$. Hence, we arrive at \eqref{eq:lyapunov}.


\subsection{Derivation of 
\texorpdfstring{\eqref{eq:modal_cross_covariance}}{modal cross covariance}}
\label{app:modal_cross_covariance}

Using \eqref{eq:modal_stochastic_response}, the It\^{o} isometry, and
$\bfb Q=\bfb B_0\bfb B_0^T$, the nonconjugate cross-covariance between $\mresp_i(t)$ and $\mresp_j(t)$ is
\begin{align*}
\mathbb{E}
\left[
\mresp_i(t)\mresp_j^T(t)
\right]
&=
\int_0^t
e^{\lambda_i(t-\tau)}
e^{\lambda_j(t-\tau)}
{\reigv}_i
{\leigv}_i
\bfb Q
{\leigv}_j^T
{\reigv}_j^T
\,d\tau
\\
&=
{\reigv}_i
\left(
{\leigv}_i\bfb Q{\leigv}_j^T
\right)
{\reigv}_j^T
\frac{
e^{(\lambda_i+\lambda_j)t}-1
}{
\lambda_i+\lambda_j
} \; .
\end{align*}

For a stable system,
$\operatorname{Re}\{\lambda_i+\lambda_j\}<0$,
and thus $\lim_{t\rightarrow\infty}
e^{(\lambda_i+\lambda_j)t}
=0$.
Hence, in the stationary limit,
\begin{equation}
\lim_{t\rightarrow\infty}
\mathbb{E}
\left[
\mresp_i(t)
\mresp_j^T(t)
\right]
=
-\frac{
{\reigv}_i
\left(
{\leigv}_i\bfb Q{\leigv}_j^T
\right)
{\reigv}_j^T
}{
\lambda_i+\lambda_j
} \; .
\end{equation}

Using \eqref{eq:modal_pair_covariance}, we arrive at 
\eqref{eq:modal_cross_covariance}.

\bibliographystyle{IEEEtran}
\bibliography{references}

@article{interval2022,
title = {Small-signal stability analysis of uncertain power systems via interval analysis},
journal = {Electric Power Systems Research},
volume = {212},
pages = {108339},
year = {2022},
issn = {0378-7796},
author = {Antonio Pepiciello and Alfredo Vaccaro},
}

@INPROCEEDINGS{brunogm_2026,
  author={Pinheiro, Bruno and Dotta, Daniel},
  booktitle={2026 IEEE Power \& Energy Society General Meeting (PESGM)}, 
  title={Dynamic Connectivity and Local Frequency Strength under Stochastic Variations}, 
  year={2026},
  volume={},
  number={},
  pages={1-5},
  doi={10.1109/PESGM58988.2026.11693538}}

@ARTICLE{Yuqi2026,
  author={Xu, Yuqi and Wan, Can and He, Zihan and Liu, Hui and Ju, Ping},
  journal={IEEE Transactions on Power Systems}, 
  title={Stochastic Stability Oriented Frequency Control: A Probabilistic Forecasting-Based High-Order Stochastic {K}oopman Approach}, 
  year={2026},
  volume={41},
  number={5},
  pages={3323-3335},
  doi={10.1109/TPWRS.2026.3674119}}

@ARTICLE{Hafsah2025,
  author={Mushtaq, Hafsah and Mir, Abdul Saleem and Senroy, Nilanjan},
  journal={IEEE Transactions on Power Systems}, 
  title={Impact of Auto-Correlation of Wind Speed on the Dynamics of Grid-Connected Doubly Fed Induction Generator}, 
  year={2025},
  volume={40},
  number={4},
  pages={3427-3438},
  doi={10.1109/TPWRS.2024.3523070}}

@article{bendong2026,
title = {Stochastic dynamic stability assessment with high penetration of uncertain sources: Uncertainty modeling, solutions and applications},
journal = {Cyber-Phys. Energy Syst.},
volume = {2},
number = {1},
pages = {13-24},
year = {2026},
issn = {3050-7448},
doi = {https://doi.org/10.1016/j.cpes.2025.11.001},
author = {Bendong Tan and Junbo Zhao}
}

@article{bruno_sin,
title = {{Evaluation of the impact of {IBR} on the frequency dynamics in the Brazilian power system}},
journal = {Sustainable Energy, Grids and Networks},
volume = {43},
pages = {101827},
year = {2025},
issn = {2352-4677},
author = {Bruno Pinheiro and M.C. {Llerena Velasquez} and Giovane Faria and A.F.C. Aquino and Diego Issicaba and Daniel Dotta},
}

@ARTICLE{CYChung_2018,
  author={Zhou, Jingsen and Ke, Deping and Chung, C. Y. and Sun, Yuanzhang},
  journal={IEEE Transactions on Power Systems}, 
  title={A Computationally Efficient Method to Design Probabilistically Robust Wide-Area {PSSs} for Damping Inter-Area Oscillations in Wind-Integrated Power Systems}, 
  year={2018},
  volume={33},
  number={5},
  pages={5692-5703},
  doi={10.1109/TPWRS.2018.2815534}}

@ARTICLE{murilo2020,
  author={Bento, Murilo E. C.},
  journal={IEEE Transactions on Power Systems}, 
  title={Fixed Low-Order Wide-Area Damping Controller Considering Time Delays and Power System Operation Uncertainties}, 
  year={2020},
  volume={35},
  number={5},
  pages={3918-3926},
  doi={10.1109/TPWRS.2020.2978426}}

@ARTICLE{CYChung2012,
  author={Wong, Wah Cheung and Chung, C. Y.},
  journal={IEEE Transactions on Power Systems}, 
  title={Coordinated Damping Control Design for {DFIG}-Based Wind Generation Considering Power Output Variation}, 
  year={2012},
  volume={27},
  number={4},
  pages={1916-1925},
  doi={10.1109/TPWRS.2012.2190110}}

@article{rev_milanovic2019,
title = {Existing approaches and trends in uncertainty modelling and probabilistic stability analysis of power systems with renewable generation},
journal = {Renewable and Sustainable Energy Reviews},
volume = {101},
pages = {168-180},
year = {2019},
issn = {1364-0321},
doi = {https://doi.org/10.1016/j.rser.2018.10.027},
author = {Kazi Nazmul Hasan and Robin Preece and Jovica V. Milanović},
}

@ARTICLE{rongcan2026,
  author={Xiao, Rongcan and Wan, Can and He, Zhiqiang and Ju, Ping},
  journal={IEEE Transactions on Power Systems}, 
  title={Probabilistic Small-Signal Stability Analysis of Power Systems With Renewable Energy: A Truncated {GMM}-Based Analytical Method}, 
  year={2026},
  volume={41},
  number={2},
  pages={1294-1307},
  doi={10.1109/TPWRS.2025.3608285}}

@ARTICLE{luke2026,
  author={Benedetti, Luke Ian and Egea-Àlvarez, Agustí and Preece, Robin and Papadopoulos, Panagiotis N.},
  journal={IEEE Transactions on Power Systems}, 
  title={Enabling Characterisation of Dynamic Interactions With Probabilistic Small-Signal Analysis in Converter-Integrated Power Systems}, 
  year={2026},
  volume={41},
  number={2},
  pages={942-954},
  doi={10.1109/TPWRS.2025.3612972}}

@ARTICLE{ignacio_2025,
  author={Ponce, Ignacio and Milano, Federico},
  journal={IEEE Transactions on Power Systems}, 
  title={Local Synchronization of Power System Devices}, 
  year={2025},
  volume={40},
  number={5},
  pages={4194-4204},
  doi={10.1109/TPWRS.2025.3543314}}

@article{iskakov2023,
title = {Estimation of the location of inter-area oscillations and their interactions in electrical power systems using {L}yapunov modal analysis},
journal = {International Journal of Electrical Power \& Energy Systems},
volume = {153},
pages = {109374},
year = {2023},
issn = {0142-0615},
author = {Alexey B. Iskakov and Evgeniy Yu. Kutyakov and Nikita V. Tomin and Daniil A. Panasetsky and Alexander N. Abramenkov and Sergey V. Dushin},
}

@book{kloeden1999numerical,
  author    = {Peter E. Kloeden and Eckhard Platen},
  title     = {Numerical Solution of Stochastic Differential Equations},
  edition   = {3rd},
  publisher = {Springer},
  year      = {1999}
}

@article{luke_SEGN,
title = {Exploring the variation of critical modal properties in power systems with converter-interfaced units},
journal = {Sustainable Energy, Grids and Networks},
volume = {43},
pages = {101841},
year = {2025},
issn = {2352-4677},
doi = {https://doi.org/10.1016/j.segan.2025.101841},
author = {Luke Ian Benedetti and Panagiotis N. Papadopoulos and Robin Preece}
}

@ARTICLE{taulant_2026,
  author={Kërçi, Taulant and Milano, Federico},
  journal={IEEE Transactions on Power Systems}, 
  title={Frequency Quality Metrics Based on Second-Order Derivative and Autocorrelation}, 
  year={2026},
  volume={41},
  number={5},
  pages={4045-4048},
  doi={10.1109/TPWRS.2026.3707412}}

@ARTICLE{Samuel2019,
  author={Chevalier, Samuel C. and Hines, Paul D. H.},
  journal={IEEE Transactions on Power Systems}, 
  title={Mitigating the Risk of Voltage Collapse Using Statistical Measures From {PMU} Data}, 
  year={2019},
  volume={34},
  number={1},
  pages={120-128},
  doi={10.1109/TPWRS.2018.2866484}}

@ARTICLE{georgia_2020,
  author={Pierrou, Georgia and Wang, Xiaozhe},
  journal={IEEE Transactions on Circuits and Systems I: Regular Papers}, 
  title={Analytical Study of the Impacts of Stochastic Load Fluctuation on the Dynamic Voltage Stability Margin Using Bifurcation Theory}, 
  year={2020},
  volume={67},
  number={4},
  pages={1286-1295},
  doi={10.1109/TCSI.2019.2943509}}

@article{2019_kamwa,
title = {Fundamental study of common mode small-signal frequency oscillations in power systems},
journal = {International Journal of Electrical Power \& Energy Systems},
volume = {106},
pages = {201-209},
year = {2019},
issn = {0142-0615},
doi = {https://doi.org/10.1016/j.ijepes.2018.09.042},
author = {Ruichao Xie and Innocent Kamwa and Dmitry Rimorov and Ali Moeini}
}

@ARTICLE{adeen2023,
  author={{M. Adeen et al.}},
  journal={IEEE Trans. on Power Syst.}, 
  title={On the Calculation of the Variance of Algebraic Variables in Power System Dynamic Models With Stochastic Processes}, 
  year={2023},
  volume={38},
  number={2},
  pages={1739-1742},
  doi={10.1109/TPWRS.2022.3226076}}

@ARTICLE{Goodarz2016,
  author={Ghanavati, Goodarz and Hines, Paul D. H. and Lakoba, Taras I.},
  journal={IEEE Transactions on Power Systems}, 
  title={Identifying Useful Statistical Indicators of Proximity to Instability in Stochastic Power Systems}, 
  year={2016},
  volume={31},
  number={2},
  pages={1360-1368},
  doi={10.1109/TPWRS.2015.2412115}}

@incollection{Benner1999SLICOT,
  author    = {Benner, Peter and Mehrmann, Volker and Sima, Vasile and Van Huffel, Sabine and Varga, Andras},
  title     = {SLICOT -- A Subroutine Library in Systems and Control Theory},
  booktitle = {Applied and Computational Control, Signals, and Circuits},
  editor    = {Datta, Biswa Nath},
  volume    = {1},
  chapter   = {10},
  pages     = {499--539},
  publisher = {Birkh{\"a}user},
  address   = {Boston, MA},
  year      = {1999},
  doi       = {10.1007/978-1-4612-0571-5_10}
}

@ARTICLE{Hongyu2019,
  author={{Hongyu Li et al.}},
  journal={IEEE Trans. on Power Syst.}, 
  title={Analytic Analysis for Dynamic System Frequency in Power Systems Under Uncertain Variability}, 
  year={2019},
  volume={34},
  number={2},
  pages={982-993},
  doi={10.1109/TPWRS.2018.2873410}}

@article{iskakov2021,
title = {Lyapunov modal analysis and participation factors applied to small-signal stability of power systems},
journal = {Automatica},
volume = {132},
pages = {109814},
year = {2021},
issn = {0005-1098},
doi = {https://doi.org/10.1016/j.automatica.2021.109814},
author = {Alexey B. Iskakov and Igor B. Yadykin}
}

@ARTICLE{federico2013,
  author={Milano, Federico and Zárate-Miñano, Rafael},
  journal={IEEE Transactions on Power Systems}, 
  title={A Systematic Method to Model Power Systems as Stochastic Differential Algebraic Equations}, 
  year={2013},
  volume={28},
  number={4},
  pages={4537-4544},
  doi={10.1109/TPWRS.2013.2266441}}

@article{milano2022power,
  title={Power system modelling as stochastic functional hybrid differential-algebraic equations},
  author={Milano, Federico and Liu, Muyang and Murad, Mohammed AA and J{\'o}nsd{\'o}ttir, Gu{\dh}r{\'u}n M and Tzounas, Georgios and Adeen, Muhammad and Ortega, {\'A}lvaro and Dassios, Ioannis},
  journal={IET Smart Grid},
  volume={5},
  number={5},
  pages={309--331},
  year={2022},
  publisher={Wiley Online Library}
}

@ARTICLE{vorobev_2019,
  author={Vorobev, Petr and Greenwood, David M. and Bell, John H. and Bialek, Janusz W. and Taylor, Philip C. and Turitsyn, Konstantin},
  journal={IEEE Transactions on Power Systems}, 
  title={Deadbands, Droop, and Inertia Impact on Power System Frequency Distribution}, 
  year={2019},
  volume={34},
  number={4},
  pages={3098-3108},
  doi={10.1109/TPWRS.2019.2895547}}

@ARTICLE{chow2016,
  author={Wilches-Bernal, Felipe and Chow, Joe H. and Sanchez-Gasca, Juan J.},
  journal={IEEE Transactions on Power Systems}, 
  title={A Fundamental Study of Applying Wind Turbines for Power System Frequency Control}, 
  year={2016},
  volume={31},
  number={2},
  pages={1496-1505},
  doi={10.1109/TPWRS.2015.2433932}}

@article{bouterakos2025,
title = {On the eigenvalue tracking of large-scale systems},
journal = {International Journal of Electrical Power \& Energy Systems},
volume = {172},
pages = {111310},
year = {2025},
issn = {0142-0615},
author = {Andreas Bouterakos and Joseph McKeon and Georgios Tzounas}
}

@article{tzounas2019modal,
  title={Modal participation factors of algebraic variables},
  author={G. Tzounas and Dassios, Ioannis and Milano, Federico},
  journal={IEEE Transactions~on Power Systems},
  volume={35},
  number={1},
  pages={742--750},
  year={2019},
  publisher={IEEE}
}

@article{benedetti2024modal,
  title={A modal contribution metric for quantifying small-signal variability in power systems with converter-interfaced generation},
  author={Benedetti, Luke and Papadopoulos, Panagiotis N and Egea-{\`A}lvarez, Agust{\'\i}},
  journal={IEEE Transactions on Power Systems},
  volume={40},
  number={3},
  pages={2636--2647},
  year={2024},
  publisher={IEEE}
}


\newpage

\begin{biography}[{\includegraphics[width=1in,height=1.25in,clip,keepaspectratio]{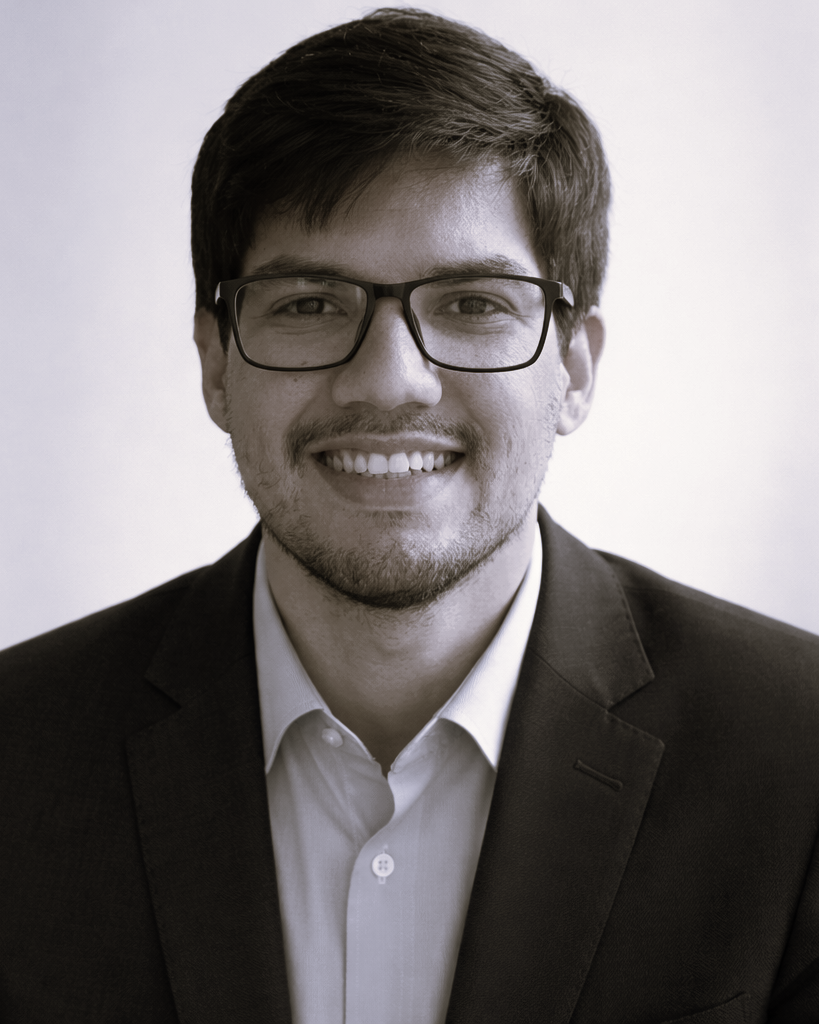}}]  {Bruno Pinheiro} (M'26) received the B.S. degree in electrical engineering from the Federal University of Amazonas, Brazil, in 2018, and the M.Sc. and Ph.D. degrees in electrical engineering from the University of Campinas (UNICAMP), Brazil, in 2021 and 2025, respectively. In Apr.-Oct. 2024, he was
a visiting researcher at  University College Dublin
(UCD), Ireland. From Dec.~2025 to May~2026, he was a postdoctoral researcher with UNICAMP. Since May 2026, he has been a Postdoctoral Researcher with UCD, Dublin, Ireland. His research interests include power system stability, control, and modeling.
\end{biography}

\vspace{-5mm}

\begin{biography}[{\includegraphics[width=1in,height=1.25in,clip,keepaspectratio]{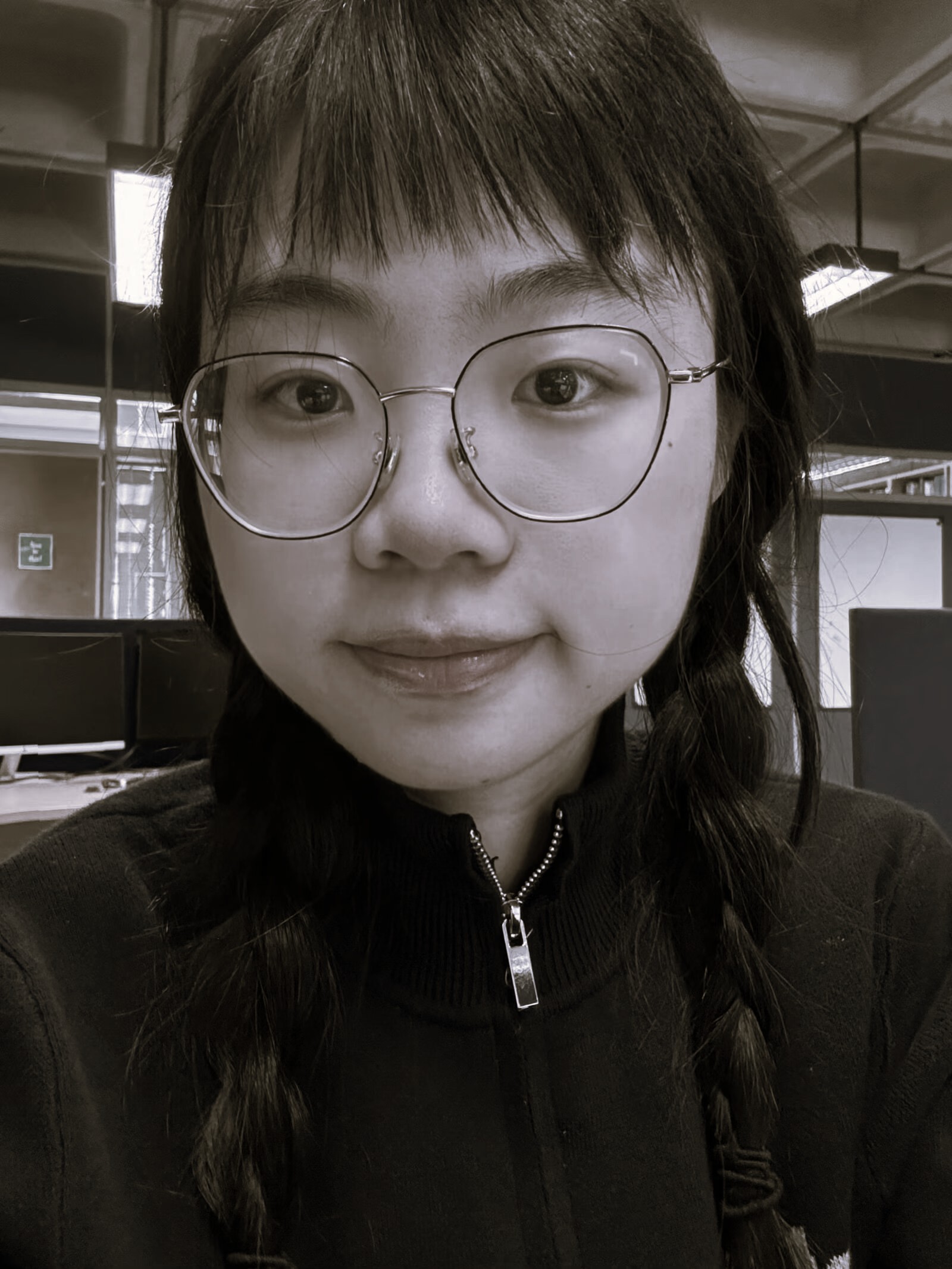}}]
{Liya Huang} (S'26) received the B.S. degree in electrical engineering and automation from Beijing Institute of Technology, Beijing, China, and the M.S. degree in electrical engineering from Central South University, Changsha, China. Since 2024, she has been pursuing the Ph.D. degree in electrical engineering with University College Dublin (UCD), Dublin, Ireland. Her research interests include power system stability and simulation.
\end{biography}

\vspace{-5mm}

\begin{biography}[{\includegraphics[width=1in,height=1.25in,clip,keepaspectratio]{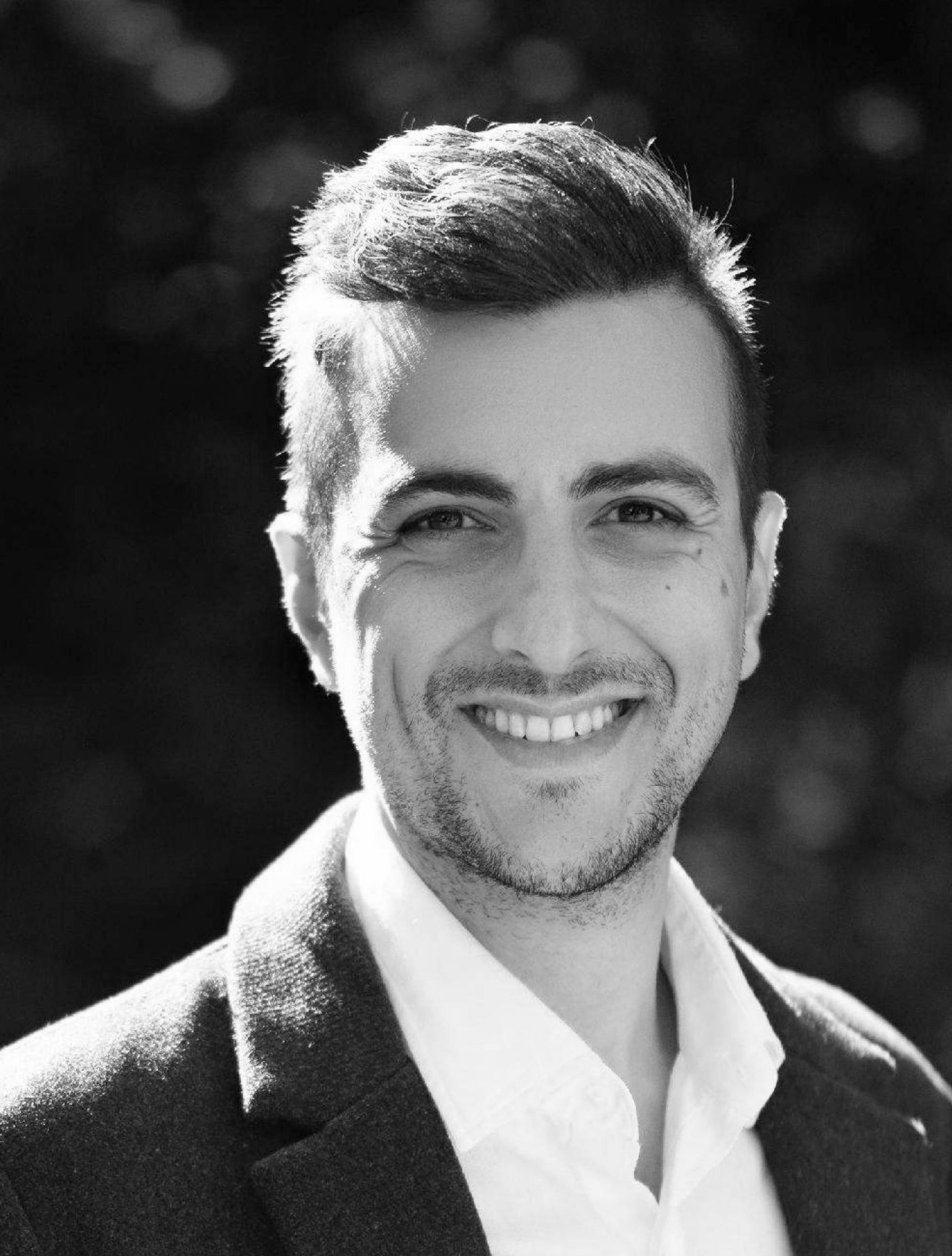}}]  {Georgios Tzounas} (SM'26) received the Diploma (M.E.) in Electrical and Computer Engineering from the National Technical University of Athens, Greece, in 2017, and the Ph.D.~from University College Dublin (UCD), Ireland, in 2021.  In Jan.-Apr.~2020, he was a visiting researcher at Northeastern Univ., Boston, MA.  From Oct.~2020 to Apr.~2023, he was a postdoctoral researcher with UCD (2020-2022) and ETH Z\"urich (2022-2023).  Since Apr.~2023, he has been an Assistant Professor with the School of Electrical and Electronic Engineering at UCD.  His primary research area is power system dynamics.
\end{biography}


\vfill

\end{document}